\documentclass[a4paper,10pt]{article}
\usepackage{authblk}
\usepackage[utf8]{inputenc}
\usepackage{moreverb,url}
\usepackage{booktabs}
\usepackage[colorlinks,bookmarksopen,bookmarksnumbered,citecolor=red,urlcolor=red]{hyperref}
\usepackage[algo2e,algonl,ruled]{algorithm2e}
\usepackage{tcolorbox,graphicx,subfig,listings,pdfpages}
\usepackage{amsmath}
\definecolor{rikenred}{cmyk}{0.0, 0.802, 0.772, 0.345}
\newcommand\BibTeX{{\rmfamily B\kern-.05em \textsc{i\kern-.025em b}\kern-.08em
T\kern-.1667em\lower.7ex\hbox{E}\kern-.125emX}}

\providecommand{\keywords}[1]{\textbf{\textit{Index terms---}} #1}

\title{CUBE: A scalable framework for large-scale industrial simulations}

\author[1,*]{Niclas Jansson}
\author[1,*]{Rahul Bale}
\author[1]{Keiji Onishi}
\author[1,2]{Makoto Tsubokura}
\affil[1]{RIKEN Advanced Institute for Computational Science, Kobe, Japan.}
\affil[2]{Department of Computational Science, Graduate School of System Informatics, \\ Kobe University, Kobe, Japan.}
\affil[*]{Corresponding aurthors: leifniclas.jansson@riken.jp, rahul.bale@riken.jp}
\date{}
\begin{document}

\maketitle

\begin{abstract}
Writing high performance solvers for engineering applications is a delicate task. These codes are often developed on an application to application basis, highly optimized to solve a certain problem. Here, we present our work on developing a general simulation framework for efficient computation of time resolved approximations of complex industrial flow problems - Complex Unified Building cubE method (\textsc{Cube}). To address the challenges of emerging, modern supercomputers, suitable data structures and communication patterns are developed and incorporated into \textsc{Cube}. We use a Cartesian grid together with various immersed boundary methods to accurately capture moving, complex geometries. The asymmetric workload of the immersed boundary is balanced by a predictive dynamic load balancer, and a multithreaded halo-exchange algorithm is employed to efficiently overlap communication with computations. Our work also concerns efficient methods for handling the large amount of data produced by large-scale flow simulations, such as scalable parallel I/O, data compression and in-situ processing.
\end{abstract}

\section{Introduction}
In the past decades, rapid advances in computer architecture and parallel computing, enabling faster and more accurate numerical simulations, have positioned computational fluid dynamics (CFD) as a standard tool in many areas of science and engineering, capable of competing with full-scale wind tunnel experiments. However, despite the remarkable progress in computational power, accurate engineering simulations are still very time consuming and pose several challenges with respect to mesh generation and numerical approximation, in particular in the context of high performance computing (HPC).

CFD simulations of flow phenomena observed in industrial applications, such as aerodynamics of vehicles and aircrafts, flow around wind turbines, etc., require high computational power. Consequently, large simulations of practical interest are not feasible without parallel computing. Parallelization can significantly alleviate the computational burden of such large simulations by distributing the work across many processors. In parallel computing, parallel efficiency and scalability of a numerical method or a flow solver determine how efficiently the computing resources are being utilized. The chosen meshing technique, representation of geometries and numerical methods not only have an impact on the overall accuracy of the simulation, they greatly affect performance and parallel scalability of a solver. This is even more important in today's extreme scale computing environment where simulation frameworks must be capable of harnessing the power of tens of thousands of cores. At this scale, the parallelization strategy of a solver is of uttermost importance, even the slightest amount of workload imbalance, poor communication patterns or serial sections can greatly affect scalability.

In this paper, we present \textsc{Cube}, a framework for efficient computation of time resolved approximations of complex industrial flow problems with moving geometries. The framework is based on a block structured Cartesian meshing technique called Building Cube Method (BCM) \cite{naka03}. The BCM framework enables easy parallelization and formulation of efficient numerical kernels. A constraint based immersed boundary (IB) method is used to model the flow around complex geometries encountered in industrial applications \cite{shir09,bhal13,pata00}.  A hybrid MPI+OpenMP based parallelization is used to hide communication cost by overlapped communication patterns, together with a multi-constraint based load balancing framework to enable scalable simulations. Throughout the paper, we will mainly focus on the design and parallelization aspects of the framework.

There are two main approaches that can be categorized under IB methods, namely continuous forcing approach and the discrete forcing approach \cite{mitt05}. The categorization is based on whether the forcing due to an immersed body is applied before or after discretization of the governing fluid flow equations. As the names imply, in the continuous and discrete forcing approaches the forcing is applied before and after the discretization of the governing equations, respectively. An appealing feature of the continuous forcing methods is that they are independent of the discretization employed \cite{pesk01,shir09, bhal13,glow99,pata00}. The discrete forcing methods on the other hand are discretization dependent, but these methods are capable of sharply representing the  immersed body interface \cite{mitt08,onis13,tsen03,fadl00}.

In the present work we use a continuous forcing technique, specifically the constraint based IB method, developed previously by Patankar and co-workers \cite{pata00,shir09,bhal13}, that enables a solver to handle complex immersed bodies without any special treatment during pre-processing. For simulations involving industrial applications, minimization of the overall pre-processing time is important.  The overall reduction in pre-processing time depends on whether the IB method employed allows for ``non-water-tight'' immersed body geometries. A water-tight geometry is one in which the volume inside the  immersed body is separated from the volume outside by a closed surface. Industrial scale or production grade CAD data are not always water-tight. Therefore, methods that require water-tight geometries necessitate  pre-processing. Thus, we choose the constraint based IB method as it allows non-water-tight geometries, reducing or eliminating the pre-processing of CAD data. Furthermore, the continuous forcing technique coupled with a Lagrangian representation of the immersed geometry is more robust and versatile in handling complex moving geometries compared to discrete forcing methods. A limitation of continuous forcing approach is its inability to sharply resolve fluid-immersed body interfaces. This limitation can be addressed through higher spatial resolution near the fluid-immersed body interface. Thus, we couple the immersed boundary method with BCM so as to achieve high spatial resolutions close to  immersed body.

Load balancing is a key factor that strongly influences the efficiency of a parallel code. In a massively parallel computing environment even the slightest workload imbalance can severely affect the performance of a code. Thus, load balancing is an essential aspect of a large scale simulations. It is closely linked to the fundamental aspect of parallel computing namely data decomposition. Data decomposition is typically done either through a pre-processor before the main simulation, or during the initial steps of the simulation.  The decomposition is often carried out on the underlying computational mesh that discretizes the computational domain. The aim of such decomposition is to equally distribute the unit of discretization such as tetrahedral or hexahedral cells, blocks across all the workers. Most decomposition techniques assume that the workload of each discretization unit is equal. Although this is true for many applications, there are a large class of problems for which it is not.  Lagrangian-Eulerian based immersed boundary methods are one such class of problems. In Lagrangian-Eulerian approach the immersed body is represented by a discrete set of Lagrangian particles which are free to move relative to the fixed background Eulerian mesh. For such a framework, the computational cost of cells or cubes that overlap with Lagrangian particles is different from that of non-overlapping cells. In this work we develop a general multi constraint load balancing technique  based on the intelligent remapping approach PLUM \cite{Oliker1998}, to decompose the Lagrangian-Eulerian system.

The outline of the paper is the following; In Section \ref{sec:math} an overview of the numerical methods is given. Our general framework is presented in Section \ref{sec:framework} and paralleization, load balancing and I/O strategies in Section \ref{sec:para} together with performance analysis and parallel scalability in Section \ref{sec:scala}. We present a typical industrial application in Section \ref{sec:IA}, give conclusions and outline future work in Section \ref{sec:sum}.

%The outline of the paper is the following; First an overview of the numerical methods is given. Our general framework is then presented together with parallelization, load balancing and I/O strategies. A performance analysis and parallel scalability is presented. Finally a typical industrial application is presented together with conclusions and an outline of future work.

%together with parallelization, load balancing and I/O strategies that enable large scale simulations. 

%an overview of our unified simulation framework is given and various design decisions are discussed. Our parallelization strategy is presented in Section \ref{sec:para}. In Section \ref{sec:loadb} and Section \ref{sec:iostart} we present our load balancing and I/O strategies that enable large scale simulations. We give conclusions and outline future work in Sections \ref{sec:app} $\&$ \ref{sec:sum}.

\section{Mathematical Framework}
\label{sec:math}
 We consider a viscous, incompressible fluid with immersed bodies. The governing equations for such a setup are given by
 \begin{align}
   & \rho\left( \dfrac{\partial \mathbf{u}}{\partial t}  + (\mathbf{u} \cdot \nabla) \mathbf{u} \right)=-\nabla p + \mu \nabla^{2} \mathbf{u} + \mathbf{f}  & \textrm{in} \;\; \Omega, \; \label{eqn-gNS}\\
   & \nabla \cdot \mathbf{u}=0   & \textrm{in} \;\; \Omega, \; \label{eqn-gcont} \\
   & \dfrac{\partial \mathbf{X}}{\partial t} = \mathbf{U}  & \textrm{in} \;\; \Psi, \; \label{eqn-lagvel} \\
%   & \mathbf{U} = \mathcal{I}[\mathbf{u}] \label{eqn-lagu}  & \textrm{in} \;\; \Psi, \; \\
   & \mathbf{f} = \mathcal{P}[\mathbf{F}] & \textrm{in} \;\; \Omega_s, \label{eqn-constrantf}
  \end{align}
where $\mathbf{u}$ is the velocity field, $\rho$ is the density, $p$ is the pressure, and $\mu$ is the dynamic viscosity of the fluid. The physical space occupied by the fluid along with the  immersed body is denoted by $\Omega$. The domain $\Omega$ is divided into $\Omega_{f}$ and $\Omega_{s}$ to represent the fluid and the  immersed body domains, respectively, $\Omega = \Omega_s \cup \Omega_f$. The  immersed body, fluid interface is denoted by $\partial\Omega_{sf}$. Note that bold face variables, such as $\mathbf{u}$, are vector quantities, and non-bold faced quantities, such as $p$, are scalar quantities.

A material description is used for immersed body representation, in which the structure resides in a Lagrangian domain $\Psi$. The physical coordinates of the immersed body, $\mathbf{X(c,}t)$, is updated with the immersed body velocity $\mathbf{U(c,}t)$, which is defined on the Lagrangian domain $\Psi$. Here, $\mathbf{c}$ is the material coordinate of the immersed body in the Lagrangian domain, $\mathbf{c}\in \Psi$, and the physical coordinate, $\mathbf{X}$, of each material point is defined in $\Omega$, $\mathbf{X(c,}t) \in \Omega$. A material quantity $\mathbf{V}$ is related to its Eulerian counterpart $\mathbf{v}$ by $\mathbf{V} = \mathcal{I}[\mathbf{v}]$ The operator $\mathcal{I}$  maps $\mathbf{v}$ from $\Omega$ to $\mathbf{V}$ in the Lagrangian domain $\Psi$.  A detailed discussion of the Lagrangian-Eulerian framework is presented in Section. \ref{sec-lageul}.

Presence of the immersed body in the fluid is modeled by the body force $\mathbf{f}$ in the momentum equation. The body force arises due to the rigidity constraints imposed by the immersed body in $\Omega_{s}$, thus by definition $\mathbf{f}$ is non-zero in $\Omega_{s}$ and zero in $\Omega_{f}$. The constraint force due to the immersed body is evaluated in the Lagrangian domain, defined by $\mathbf{F}$,  before it is mapped onto its Eulerian counterpart $\mathbf{f}$. The mapping operator $\mathcal{P}$ maps a Lagrangian variable onto its Eulerian counterpart. The operators $\mathcal{P}$ and $\mathcal{I}$  are defined and discussed in the following section.

The rigidity constraint is imposed in the immersed body domain $\Omega_s$ and at fluid immersed body interface $\partial\Omega_{sf}$ as
\begin{align}
  & \nabla \cdot \mathbf{D}(\mathbf{u} - \mathbf{u}_{s} ) = 0 \;\;\;\;\; \textrm{in} \;\; \Omega_{s}, \label{eqn-constraint1}\\
  & \mathbf{D}(\mathbf{u} - \mathbf{u}_{s} ) \cdot \hat{\mathbf{n}} = 0 \;\;\;\;\; \textrm{on} \;\; \partial\Omega_{sf},  \label{eqn-constraint2}
\end{align}
where $\mathbf{u}_{s}$ is the specified velocity of the immersed body, and $\mathbf{D}$ is the deformation rate tensor given by
\begin{equation*}
 \mathbf{D}(\mathbf{u}) = \frac{1}{2}(\nabla\mathbf{u} + \nabla\mathbf{u}^{T}).
\end{equation*}

Eq. \eqref{eqn-constraint1} and Eq. \eqref{eqn-constraint2}  physically imply that the rate of deformation of the rigid, flow induced velocities on the immersed body velocity must be zero. In the case where motion or velocities induced by the fluid on the immersed body is not of interest, the above equations are replaced by the following
\begin{align}
  & (\mathbf{u} - \mathbf{u}_{s} ) = 0 \;\;\;\;\; \textrm{in} \;\; \Omega_{s}, \nonumber \\ %\label{eqn-constraint3}\\
  & (\mathbf{u} - \mathbf{u}_{s} ) \cdot \hat{\mathbf{n}} = 0 \;\;\;\;\; \textrm{on} \;\; \partial\Omega_{sf}.\nonumber%  \label{eqn-constraint4}
\end{align}
\cite{shir09} have shown that the rigidity constraint in Eqs. \ref{eqn-constraint1} \& \ref{eqn-constraint2} give rise to a constraint force that is given by 
\begin{equation*}
\mathbf{f} = \nabla \cdot \mathbf{D}(\lambda_{r} - 2\mu \mathbf{u}_{s}).
\end{equation*}
%  \vskip-.4cm
%  \begin{equation}
%   \mathbf{f} = \nabla \cdot \mathcal{D}(\lambda_{r} - 2\mu \mathbf{u}_{s}).
%  \end{equation}
$\lambda_{r}$, in the equation, is the Lagrange multiplier that enforces rigid motion constraint in the immersed body domain ($\Omega_{s}$). $\lambda_{r}$ is analogous to the mechanical pressure, $p$, which is also a Lagrange multiplier. $p$ is a Lagrange multiplier that imposes the incompressiblity constraint. It is to be noted that we do not directly compute the Lagrange multiplier, rather an equivalent forcing upon discretization.
% given by $\mathbf{f}^{n+1}_{i,j,k}= \rho \frac{\delta \mathbf{u}}{\delta t}$, where $\delta \mathbf{u} = (\mathbf{u}_{s})_{i,j,k}^{n+1}-\mathbf{\tilde{u}}_{i,j,k}^{n+1} $. Here, $g_{i,j,k}^{n+1}$ is the discrete version of $g$ at $\mathbf{x}(i,j,k)$ ($\mathbf{x}\in\Omega$) at time $n+1$, and $\mathbf{\tilde{u}}^{n+1}_{i,j,k}$ is the intermediate velocity of the projection method discretization when marching from time $n$ to $n+1$.

\subsection{Lagrangian--Eulerian approach}
\label{sec-lageul}
Accurate representation of intricate details, down to the Eulerian mesh resolution, of 
immersed body is key to the success of the numerical method. This is especially true when the immersed body is moving. As already introduced in the previous section, a Lagrangian-Eulerian approach is used in this work because a Lagrangian description is a very accurate method of representing complex, mobile immersed bodies. The physical space $\Omega$, $\Omega_{s}$ and boundary $\partial\Omega_{sf}$ defined in the previous section are based in the Eulerian description. For the Lagrangian description we define a coordinate system $\Psi$ on which the body resides. $\Omega_{s}$ is the  projection of Lagrangian domain $\Psi$ on to Eulerian domain $\Omega$ through the Dirac delta function operator. The operators $\mathcal{P}$ and $\mathcal{I}$ are defined by the  following equations

\begin{align}
\begin{split} \label{eqn-deltap}
 \mathbf{g(x},t)  = &\mathcal{P[\mathbf{G(c}},t)], \; \textrm{where,} \\
 & \mathcal{P[\mathbf{G(c}},t)]= \int_{\Psi}\mathbf{G(c},t)\delta(\mathbf{x}-\mathbf{X}(\mathbf{c},t)) d\mathbf{c},
\end{split}\\
\begin{split} \label{eqn-deltai}
  \mathbf{V(c},t)= & \mathcal{I[\mathbf{v(x}},t)], \; \textrm{where,} \\
& \mathcal{I[\mathbf{v(x}},t)]= \int_{\Omega_{s}}\mathbf{v(x},t)\delta(\mathbf{x}-\mathbf{X}(\mathbf{c},t)) d\mathbf{x}.
\end{split}
\end{align}
The variable $\mathbf{G}$, which is defined on the Lagrangian domain ($\Psi$), is projected on to the variable $\mathbf{g}$, which is defined on the Eulerian mesh's immersed body domain ($\Omega_{s}$). Similarly, Eulerian variable $\mathbf{v}$ is interpolated on to the Lagrangian variable $\mathbf{V}$. The coordinate $\mathbf{x}$ is defined on the Eulerian domain $\Omega$, $\mathbf{x}\in \Omega$. And, $\mathbf{c}$ is the discrete Lagrangian coordinate defined on $\Psi$. $\delta(\mathbf{x})$ is the Dirac delta function defined as $\delta(\mathbf{x}) = \delta(x_1)\delta(x_2)\delta(x_3)$ in three dimensions, where $\mathbf{x}=(x_1,x_2,x_3)$.

\begin{figure}[th]
  \centering
 {\includegraphics[width=0.9\columnwidth]{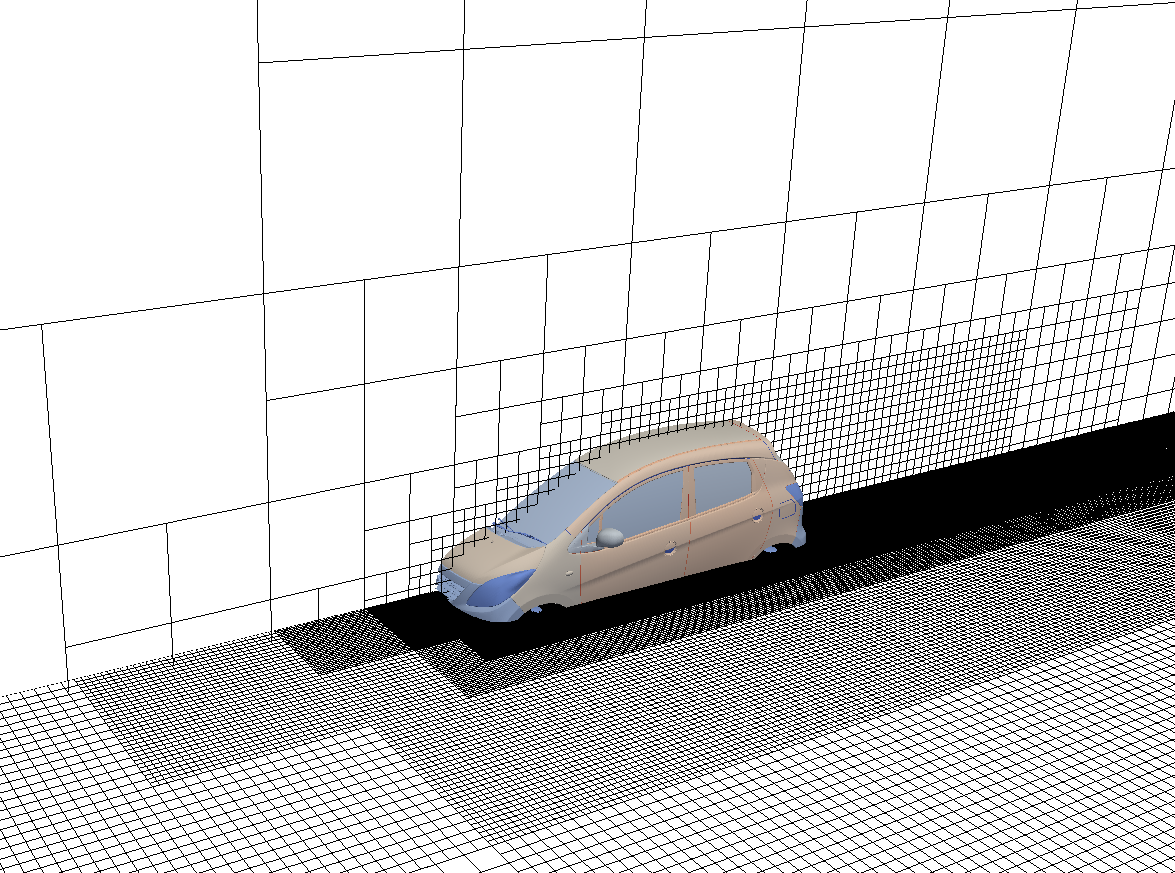}}
 \caption{\label{fig:bcm-lag-scaling} An example of a BCM mesh, the vertical plane shows the cubes and the horizontal plane shows cubes subdivided into cells.}
% \begin{subfigure}[b]{0.4\textwidth}
%   \includegraphics[width=0.4\textwidth]{cerbo/grid_cerbo_2.png}
% \end{subfigure}
% \begin{subfigure}[b]{0.4\textwidth}
%   \includegraphics[width=0.4\textwidth]{Lagrangian_scaling/WL-vs-NptsLag.pdf}
% \end{subfigure}

% \caption{(a) An example of BCM, the vertical plane shows the cubes and the horizontal plane shows cubes subdivided into cells. (b) Ratio of time spent on Lagrangian operations during a time step to the total time step on one time step plotted as a function of the ratio of number of Lagrangian points to Eulerian mesh size. }
%\label{fig:bcm-lag-scaling}
\end{figure}

% \begin{figure}[th]
%   \centering
%  {\includegraphics[width=0.4\textwidth]{Halo_exchange.pdf}}
%  \caption{\label{fig:halo-exchange} }
% \end{figure}

\begin{figure*}[t]
  \centering
 {\includegraphics[width=0.9\textwidth]{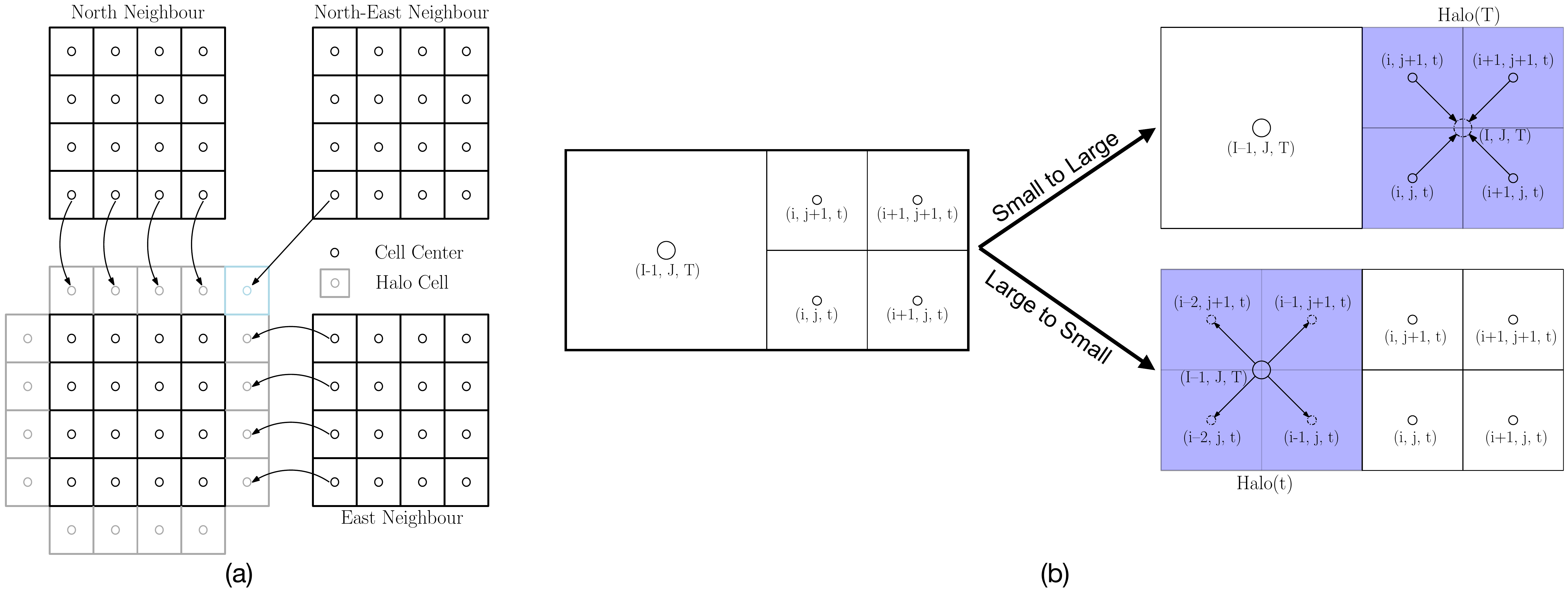}}
 \caption{\label{fig:halo-hanging-node} (a) Exchange of information between neighbouring cubes of same size. Data from neighbouring cubes are copied to adjoining halo cells. (b) Data exchange between cells of neighbouring coarse and fine cubes. In the fine to coarse data exchange the halo cell of coarse cube gets interpolated value from the interior cells of the neighbouring fine cubes. During coarse to fine data exchange, the data from the boundary cell of the coarse cube is copied into all the halo cells of the fine cube that are within the coarse cube cell.    }
\end{figure*}

\subsection{Building Cube Method}
The Eulerian domain ($\Omega$) is discretized using a structured Cartesian meshing technique called the building cube method (BCM), wherein the Eulerian domain is discretized using cubic units called cubes. The cubes are subdivided into fineer cubes in regions of interest, for example around an immersed body, to generate a set of cubes ranging from coarsest level ($l = 0$) to finest level ($l = m-1$) with a total of $l_n = m$  levels, and $l_r = m-1$ number of refinements. Although it possible in theory to represent the cubes in a heirarchical tree structure, in the present version of BCM we do not use a tree data structure to represent the cubes. A simpler, more efficient, data structure, where only the cube coordinates, size, an arbitrary index, number of cells and neighbor adjacency information is used to represent the mesh using linear arrays in this work.  An arbitrary refinement ratio can be used between a coarse and fine cube, for the results presented in this work this ratio is restricted to 2. Cubes are further subdivided into cells such that all cubes have the same number of cells.  As every cube has equal cells, all cubes are equal and independent units of work. This enables efficient and easy grid partitioning and parallelization. An example of BCM mesh is shown in Fig. \ref{fig:bcm-lag-scaling}, the vertical plane slice shows the cubes and the horizontal slice shows the cube subdivided into cells. 

A cell centered, collocated arrangement is used to discretize the governing equations on the BCM mesh, wherein the Eulerian variables ($\mathbf{u}$, $p$, and $\mathbf{f}$) are defined at the center of a cell. The coordinate of the cell center defined by the index $\mathbf{i}=(i,j,k)$, in the $t^{th}$ cube, is given by
\begin{equation*}
 \mathbf{x_i}^{t} = \mathbf{x}_{c}^{t} + \left( (i+\frac{1}{2})\Delta x_{1}^{t}, (j+\frac{1}{2})\Delta  x_{2}^{t}, (k+\frac{1}{2})\Delta  x_{3}^{t} \right),
 \label{eqn-cell-center}
\end{equation*}
where $\mathbf{x}_{c}^{t} = (x_{1c},x_{2c},x_{3c})$ is the coordinate of the base corner of a cube indexed $t$, and $\Delta\mathbf{x}^{t} = (\Delta x_{1}^{t}, \Delta x_{2}^{t}, \Delta x_{3}^{t})$ is the mesh spacing. Base corner of a cube is the corner with the minimum coordinate along the principle directions. $\mathbf{x}_{c}$ is the set of base corner coordinates of all the cubes in a mesh so that $\mathbf{x}_{c}^{t} \in \mathbf{x}_{c} $. Similarly, $\Delta\mathbf{x}^{t} \in \Delta\mathbf{x}$, where $\Delta\mathbf{x}$ is the set of mesh spacing of each cube in the BCM mesh.

Discrete stencil operations at a cube's boundary can span across into the neighbouring cube. A halo of cells from neighbouring cubes are defined around each cube to enable discrete stencil operations at cube faces (Fig. \ref{fig:halo-hanging-node}). The size of the halo is determined by the extent of the widest stencil used in the solver. For the halo exchange between cubes of same size, data is copied from the boundary interior cells of a neighbouring cubes into adjoining halo cells of a given cube. Halo exchange between coarse and fine cube involves data interpolation of data. A schematic representation of halo exchange is shown in Fig. \ref{fig:halo-hanging-node}. In case of data exchange from fine to coarse cube, the boundary interior cells are interpolated to the center of the halo cell of the coarse cube (Fig. \ref{fig:halo-hanging-node}b). The interpolation of a cell centered quantity $\phi$ from fine cube $t$ to the halo cell of coarse cube $T$ can be expressed as follows
\begin{equation*}
 \phi_{\mathbf{I}}^T = \sum_{\mathbf{p}=0}^{l_r-1} w_{\mathbf{i+p}} \phi_{\mathbf{i+p}}^t.
\end{equation*}
Here, $\mathbf{i}=(i,j)$  ($\mathbf{i}=(i,j,k)$ in 3D) is the cell index of fine cube and $\mathbf{I}$ is that of the coarse cube cell, $\mathbf{p}=(p,q)$ ($\mathbf{p}=(p,q,r)$ in 3D) is the summation index and $w$ is the interpolation weight. Data exchange from coarse cube interior boundary cell to fine cube halo cells is achieved by simply copying the coarse cell data to all the fine cells that lay within the corse cell. The coarse to fine exchange, as depicted in  Fig. \ref{fig:halo-hanging-node}, may be expressed as,
\begin{align}
 \phi_{i-1,j,t} & =\phi_{I-1,J,T} & \phi_{i-2,j,t} & =\phi_{I-1,J,T} \nonumber \\
 \phi_{i-1,j+1,t} & =\phi_{I-1,J,T} & \phi_{i-2,j+1,t} & =\phi_{I-1,J,T}. \nonumber
\end{align}
Treatment of face centered, staggered quantities is not necessary because we use a collocated arrangement in the present work.

\begin{figure}[th]
 \begin{center}
   \null\hfill
   \subfloat[Immersed body]{\label{fig:lag_mesh}\includegraphics[width=0.45\columnwidth]{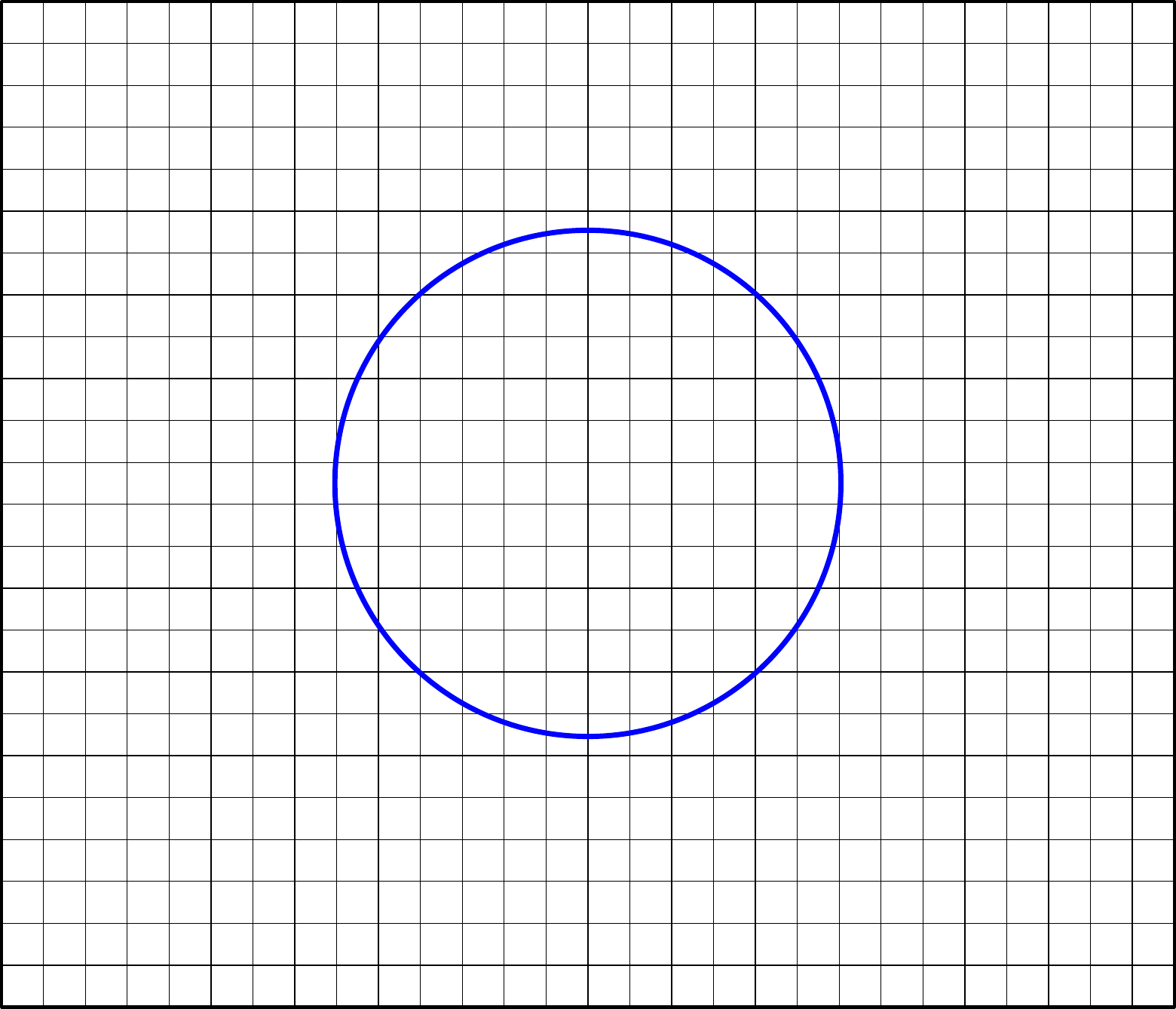}}
   \hfill
   \subfloat[Discretized into particles]{\label{fig:lag_solid}\includegraphics[width=0.45\columnwidth]{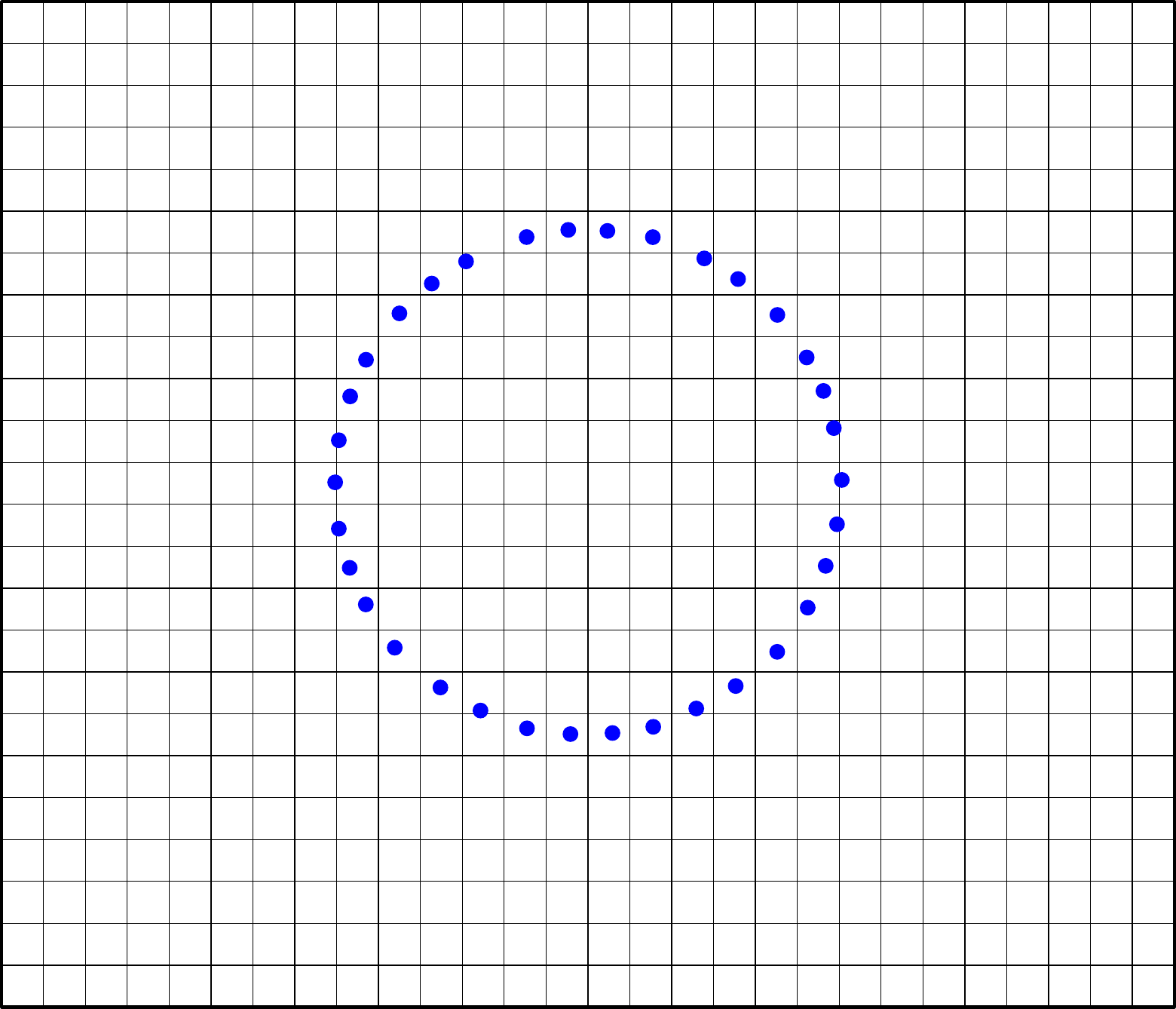}}
   \hfill\null
 \end{center}
  \caption{\label{fig:lag} Discretization of immersed body into Lagrangian particles. The surface of the cylinder in the 2D mesh is discretized into Lagrangian particles. }
\end{figure}

\begin{figure}[th]
 \begin{center}
   \null\hfill
   \subfloat[Ungrouped particles]{\label{fig:lag_seta}\includegraphics[width=0.3\columnwidth]{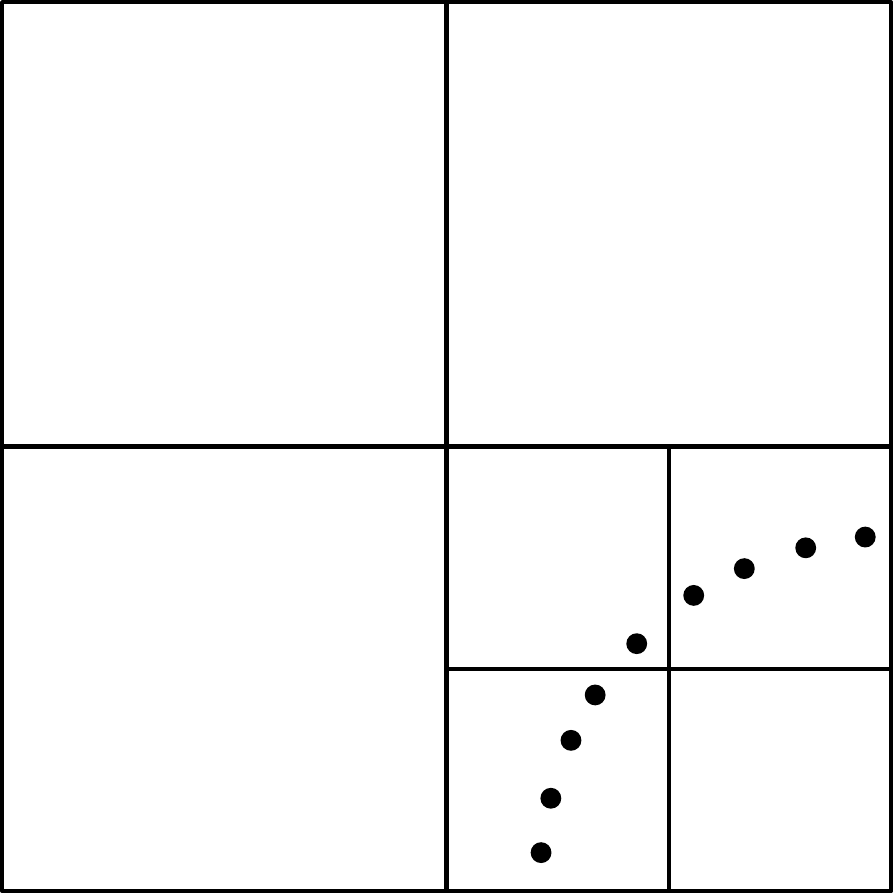}}
   \hfill
   \subfloat[Grouped particles]{\label{fig:lag_setb}\includegraphics[width=0.3\columnwidth]{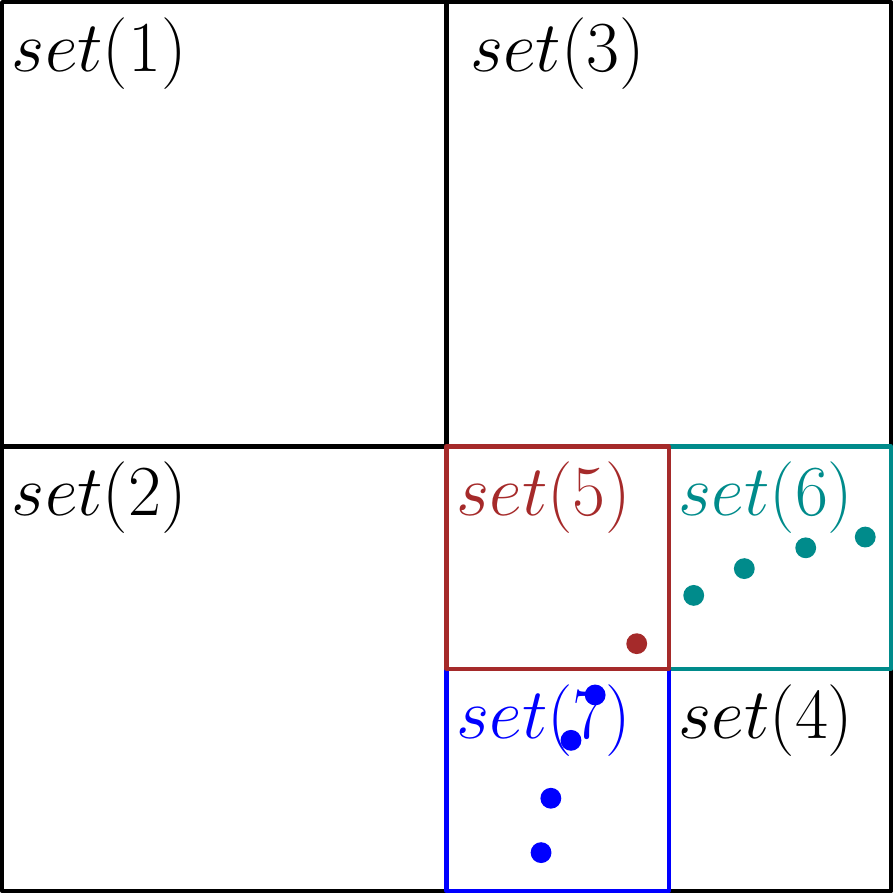}}
   \hfill
   \subfloat[Particle motion]{\label{fig:lag_setmb}\includegraphics[width=0.3\columnwidth]{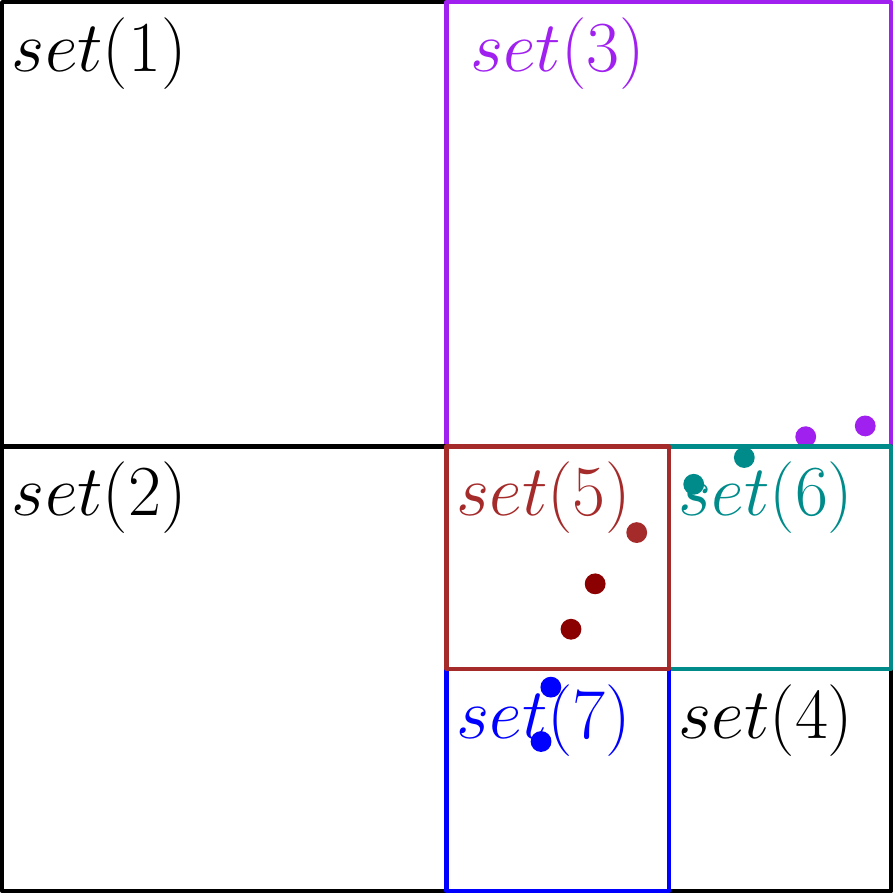}}
   \hfill\null
 \end{center}
  \caption{\label{fig:lag_set} \protect\subref{fig:lag_seta} Ungrouped Lagrangian particles and cubes of a BCM mesh. \protect\subref{fig:lag_setb} Lagrangian particles grouped into  $set$s for each cube of the BCM mesh. The $set$s are colored in black and non-black colors to identify empty and non-empty sets. \protect\subref{fig:lag_setb}  Motion of Lagrangian particles. As particles move, particles from $set(7)$ removed and added to $set(5)$. Similarly, particles from $set(6)$ are removed and added to $set(3)$.   }
\end{figure}

% \begin{figure}[th]
%  \begin{center}
%    \subfloat[Ungrouped particles]{\label{fig:lag_setma}\includegraphics[width=0.25\columnwidth]{Sphere_pts_Cube_set.pdf}}
%    \hspace*{4mm}
%    \subfloat[Grouped particles]{\label{fig:lag_setmb}\includegraphics[width=0.25\columnwidth]{Sphere_pts_Cube_set_advect.pdf}}
%  \end{center}
%   \caption{\label{fig:lag_setm} Motion of Lagrangian particles. As particles move, particles from $set(7)$ removed and added to $set(5)$. Similarly, particles from $set(6)$ are removed and added to $set(3)$.   }
% \end{figure}

\subsection{Lagrangian Data Structure}
\label{sec:bcl}
The immersed body is discretized into material/Lagrangian particles that are attached to its surface which constitute the Lagrangian domain ($\Psi$). An example of the discretization of the immersed body into Lagrangian particles is shown in Fig. \ref{fig:lag}, which shows the discretization of the surface of a 2D cylinder into Lagrangian particles. Typically, discretization of immersed body surface results in one Lagrangian particle in the Eulerian cell which immersed body surface intersects.

In order to enable simple parallelization we use the underlying cube data structure (BCM cubes) of the Eulerian mesh. The Lagrangian points are grouped into an $unordered~set$ belonging to each BCM cube. Lagrangian particles, which are inside cube $t$, are inserted into $set(t)$. Cubes that do not contain any Lagrangian particles will have empty $set$s.
The grouping of Lagrangian particles into sets is demonstrated in Fig. \ref{fig:lag_set}, which shows a $set$ assigned to each cube in the mesh shown. In the figure, empty and non-empty sets are highlighted through the use for different colors. Empty sets are colored in black while each of the non-empty set is colored with a different non-black color. The spacing between Lagrangian particles in a cube indexed $t$ is given by $\Delta\mathbf{c}^t = (\Delta c_{1}^{t}, \Delta c_{2}^{t}, \Delta c_{3}^{t})$, and a given particle is identified through the index $\mathbf{{R}} = (\mathcal{R,S,T})$. The Eulerian coordinate of a Lagrangian particle is denoted by $\mathbf{X}_{\mathcal{R,S,T}}^t$.

% As the BCM mesh is made up of different sized cubes, mesh spacing of Lagrangian particles will depend on the cube in which it lies. Particles in coarser cubes will have larger mesh spacing and ones in finer cubes will have smaller mesh spacing. When particles move between cubes of different size, they will require coarsening or refining depending on the mesh spacing the destination cube. Note that in typical applications, the immersed body is located in the region of highest mesh resolution as a result of this all the Lagrangian particles will have the same mesh spacing. Also, when the immersed body is moving, the BCM mesh is generated such that motion of the immersed body is restricted to cubes with smallest mesh spacing.

The main purpose of grouping Lagrangian particles into $set$s is to enable efficient Lagrangian-Eulerian interpolation. Interpolation from  Eulerian mesh to Lagrangian points requires, for a given Lagrangian point, information of surrounding Eulerian mesh points. Without the $set$ based Lagrangian data structure, identifying the surrounding points for interpolation is very expensive. This is because the cube indexing in a BCM mesh is arbitrary. As a result, to find surrounding points for interpolation, the cube containing the Lagrangian point in question needs to be searched. Searching is an inherently slow process which is best avoided when possible. The $set$ based data structure eliminates this search operation. The Lagrangian-Eulerian interpolations boils down to algebraic expression for finding the Eulerian indicies $\mathbf{i}$ for a given cube $\mathbf{t}$. Typical Lagragian Eulerian interaction is shown in Algorithm \ref{alg:le}. 

% \begin{comment}
\begin{algorithm2e}[t]
  \SetNoFillComment
  \SetAlgoNoLine
  \SetKwProg{Fn}{Function}{:}{}
  \Fn{L-E Interact}{

      \For{each cube in BCM mesh}{
        \eIf{set is empty}{
           cycle
        }{
            $\mathbf{i}=(\mathbf{X}_{\mathcal{R,S,T}}^t - \mathbf{x}_{c}^t)/\Delta \mathbf{x}_{c}^{t}$ + 1
            
            \textbf{Call} \textit{ L-E Interpolation}(\textit{$\mathbf{i},\mathbf{X}_{\mathcal{R,S,T}}^t,\mathbf{x}_{c}^t$})
        }
        
      }
     }

\caption{\label{alg:le} Lagrangian-Eulerian mesh interaciton.}
\end{algorithm2e}
% \end{comment}

The Lagrangian data structure is cube based, i.e. a $set$ belonging to cube $t$ is independent of the $sets$ of all the other cubes. Therefore, propogation and motion of the Lagrangian particle requires careful design of the $set$ data structure. As the particles of the $set$ of a given cube $t$ move, some of the particles of the $set(t)$ may cross the bounds of the said cube (runaway particles). The runway particles no longer belong the $set(t)$, but they may be within the bounds of one of the many neighbors of cube $t$. Hence they have to removed from $set(t)$ and added to the $set$ of the neighboring cube. A schematic representation of propogation, insertion and deletion of particles is shown in Fig. \ref{fig:lag_setmb}. This deletion and insertion of particles from and into $sets$ of cubes, respectively, has to be carried out efficiently. To enable quick insertion and deletion of particles from $sets$, the $set$ is built on an integer based $hash \;table$. Each Lagrangian particle is assigned a unique global integer identifier, which acts as the key to the $hash \;table$. Global implies across cubes as well as MPI partitions.  And, the uniqueness of the identifier is necessary to avoid duplicity of particles as they are advected across cubes. The average cost of insertion, deletion, or a look up of $hash \;table$ entries is $\mathcal{O}(1)$, and the cost of the worst case is $\mathcal{O}(N)$, where $N$ is the size of the $hash \;table$. The frequency of the running into $hash \;table$ related worst case operations depends on the choice of hash function. Consequently, the choice of a hash function that minimizes hash conflicts is key to the efficiency of the $hash \;table$.

\subsection{Lagrangian-Eulerian Interaction}
With the discretization of the Eulerian and Lagrangian domain defined, the projection and interpolation operations between the two domains will be discussed next. The interpolation from Eulerian to Lagrangian domain in discrete form is given by

\begin{equation}
 \mathbf{U}_{\mathcal{R,S,T}}^{t} = \sum\limits_{i,j,k,t} \mathbf{u}_{i,j,k}^t\delta_{\Delta}(\mathbf{x}_{i,j,k}^t - \mathbf{X}_{\mathcal{R,S,T}}^t) \Delta x_{1}^t\Delta x_{2}^t \Delta x_{3}^t,
 \label{eqn-ddeltai}
\end{equation}
where $\mathbf{U}_{\mathcal{R,S,T}}^{t}$ is the interpolated velocity at Lagrangian particle indexed $\mathcal{R,S,T}$, and $\mathbf{u}_{i,j,k}^t$ is the cell centered velocity in the cell $(i,j,k)$ in cube $t$. Similarly, the discretized version of projection operator in Eq. \eqref{eqn-deltap} is given below

\begin{equation}
 \mathbf{f}_{i,j,k}^t = \sum\limits_{\mathcal{R,S,T},t} \mathbf{F}_{\mathcal{R,S,T}}^t\delta_{\Delta}(\mathbf{x}_{i,j,k}^t - \mathbf{X}_{\mathcal{R,S,T}}^t) \Delta c_{1}^t\Delta c_{2}^t\Delta c_{3}^t.
 \label{eqn-ddeltap}
\end{equation}
Here, $\delta_{\Delta}$ is the discrete form of the Dirac delta function introduced in Eqs. \ref{eqn-deltap}~$\&$\ref{eqn-deltai} and it is given by
\begin{equation*}
 \delta_{\Delta}(\mathbf{x}) = \prod\limits_{n=1}^{3} \frac{1}{\Delta x_{n}} \phi\left(\frac{x_n}{\Delta x_{n}}\right),
 \label{eqn-discrete-delta}
\end{equation*}
where $\phi\left({x}/{\Delta x}\right)/\Delta x$ is the one dimensional discrete delta function. The following smoothed 3 point function for the one dimensional function $\phi(r)$ \cite{Yang09a} is used in this work
\begin{equation*}
 \phi(r) =  \begin{cases}
  \frac{3}{4} - r^2, & |r| \le 0.5, \\
  \frac{1}{2}(\frac{9}{4} -{3|r|} + {r^2}), & 0.5 < |r| \leq 1.5,  \\
  0, & 1.5 < |r|.
  \end{cases}
\end{equation*}
The choice of the type of Dirac delta function is important from an accuracy and a computational efficiency stand point. Wider delta function are known to give better accuracy then narrower ones \cite{Grif05,pesk01,Yang09a}, but this better accuracy comes at greater computational cost. An $n$ point wide delta function will require $n^3$ operations to interpolate at one Lagrangian particle. Therefore, from a computational cost stand point it is ideal to choose a delta function that is the narrowest (a 2-point function). A balance between accuracy and computational cost is necessary. \cite{Yang09a} have shown that the 3-point wide function in the above equation provides reasonable accuary at lower computational cost compared wider delta functions.

Eqs. \ref{eqn-ddeltai} \& \ref{eqn-ddeltap} can be expressed in a more concise form as
\begin{align}
 & \mathbf{U}_{\mathbf{R}}^{t} = \mathcal{I}_{\Delta}[\mathbf{x}_{\mathbf{i}}^{t}]\mathbf{u}_{\mathbf{i}}^{t}, \\
 & \mathbf{f}_{\mathbf{i}}^t = \mathcal{P}_{\Delta} [\mathbf{X}_{\mathbf{R}}^{t}]\mathbf{F}_{\mathbf{R}}^{t},
\end{align}
respectively. Here, $\mathcal{I}_{\Delta}$ and $\mathcal{P}_{\Delta}$ are the discrete versions of the interpolation operator $\mathcal{I}$ eq. \eqref{eqn-deltai} and $\mathcal{P}$ eq. \eqref{eqn-deltap}.

\section{Numerical algorithm}
\label{sec:n-alg}
We use a modified fractional step method \cite{chor69, shir09} to solve the system of equations in Eq. \eqref{eqn-gNS} to \eqref{eqn-constrantf}. In the first sub-step of the fractional step method, a modified momentum equation is solved to obtain a non-divergence free intermediate velocity. The constraint force due to the immersed body is not imposed at this step. In this first sub-step, the equations may be discretized using any time stepping algorithm. In \textsc{Cube}, Euler, Adams-Bashforth and implicit Crank-Nicolson schemes are available. Here we describe the numerical algorithm using the Crank-Nicolson scheme.

\begin{equation}
 \rho \frac{\mathbf{\tilde{u}} - \mathbf{u}^{n} }{\Delta t} + \rho\frac{1}{2}\mathcal{A}\left[\tilde{\mathbf{u}} + \mathbf{u}^{n}  \right] = \mu\frac{1}{2}\mathcal{D}\left[\tilde{\mathbf{u}} + \mathbf{u}^{n} \right],
 \label{eq:substep1}
\end{equation}
where a superscript indicates time, $n$ corresponds to previous time step and intermediate variables of the fractional step are identified by a tilde, and $n+1$ indicates time at the end of current new time step. $\mathcal{A}$ and $\mathcal{D}$ are discrete operators of the advection and the diffusion terms, respectively.

In the next step, the immersed body force is imposed on the intermediate velocity $\mathbf{\tilde{u}}$ to obtain the second intermediate velocity $\mathbf{u^{*}}$
\begin{equation*}
\rho \frac{\mathbf{u^{*}-\tilde{u}}}{\Delta t} = \mathcal{P}_{\Delta} [\mathbf{X}]\mathbf{F}^{n},
\end{equation*}
where the immersed body force $\mathbf{F}^{n}$, evaluated as the difference between the immersed body velocity $\mathbf{U}^{n}_{s}$ and interpolated intermediate velocity $\mathcal{I}_{\Delta}[\mathbf{x}]\mathbf{\tilde{u}}$, in the Lagrangian domain $\Psi$ is given by

\begin{equation*}
\mathbf{F}^{n} = \frac{\rho}{\Delta t} \left( \mathbf{U}_{s}^{n} - \mathcal{I}_{\Delta}[\mathbf{x}]\mathbf{\tilde{u}} \right).
\end{equation*}
The non-zero divergence of the intermediate velocity, $\mathbf{\tilde{u}}$, and that introduced by the immersed body force into $\mathbf{u}^{*}$ is removed through the projection step to yield the divergence free velocity field $\mathbf{u}^{n+1}$
\begin{equation*}
 \mathbf{u}^{n+1} = \mathbf{u}^{*} - \frac{\rho}{\Delta t} \mathcal{G}p^{n+1},
\end{equation*}
where $\mathcal{G}$ is the discrete gradient operator. The pressure $p^{n+1}$ in the projection step is obtained by solving the Poisson equation
\begin{equation}
\label{eq:poisson}
 \mathcal{D} p^{n+1} = \frac{\rho}{\Delta t} \mathcal{G} \cdot \mathbf{u^{*}}.
\end{equation}

Eq. \eqref{eq:poisson} is solved with an efficient geometric multigrid solver. Implementing a multigrid solver on a Cartesian grid with discrete forcing immersed body methods (such as ghost cell IB methods \cite{mitt08,onis13}) is a complex task. In order to retain good convergence, special treatment of the immersed body at every grid level is necessary \cite{flapping}. Since we use a continuous forcing to represent the immersed body, the effect of the body is only present in the right hand side of Eq. \eqref{eq:poisson}. Thus, it will be unaffected by grid coarsening, and only present at the fine grid level. Hence, implementation of the multigrid solver is straightforward, without the need of any special treatment of the immersed bodies at different grid levels of the multigrid solver. Furthermore, we exploit the underlying cube structure, and only create coarser grid levels on a per cube basis. Thus, we can retain the same communication pattern on all grid levels.

%\begin{figure}[th]
%  \begin{center}
%  \subfloat[Building Cube mesh.]{\label{fig:mesh}\includegraphics[width=0.2\textwidth]{restriction}}
%  \hspace*{0.1\textwidth}
%  \subfloat[The dual graph.]{\label{fig:dual}\includegraphics[width=0.2\textwidth]{prolongation}}
%  \end{center}
%\end{figure}

%\rbcom{Add a discription of the pressure solvers here? We are using two different pressure solvers in this work. SOR for validation, multigrid %for load balancing analaysis.\\
 %	And, how about a discussion on the convergence of multigrid for Lagrangian vs ghost-cell methods?}

%\cite{flapping}

Finally, the position of the Lagrangian points/particles are updated by the following equation
\begin{equation*}
 \mathbf{X}^{n+1} = \mathbf{X}^{n} + \Delta t \mathbf{U}_{s}^{n+1}.
\end{equation*}

\section{Validation}

We carry out simulation of two cases to validate implementation of the numerical method.  First, we consider the standard benchmark problem of flow around a stationary sphere. Next, to validate the method for moving immersed bodies we consider flow created by an impulsively started plate in a quiescent fluid.
\subsection{Flow around a sphere}

 Flow around a sphere is a widely used case to validate three dimensional flows. There are numerous experimental \cite{clif78} and numerical studies \cite{john99,mitt08,onis13} that have investigated this flow at various Reynolds numbers against which we can validate our simulations. In the present study we carried out numerical simulations for Reynolds numbers ranging from 100 to 1000 (Reynolds number is based on sphere diameter). 
 
\begin{figure}[h]
 \centering
\includegraphics[width=0.5\textwidth]{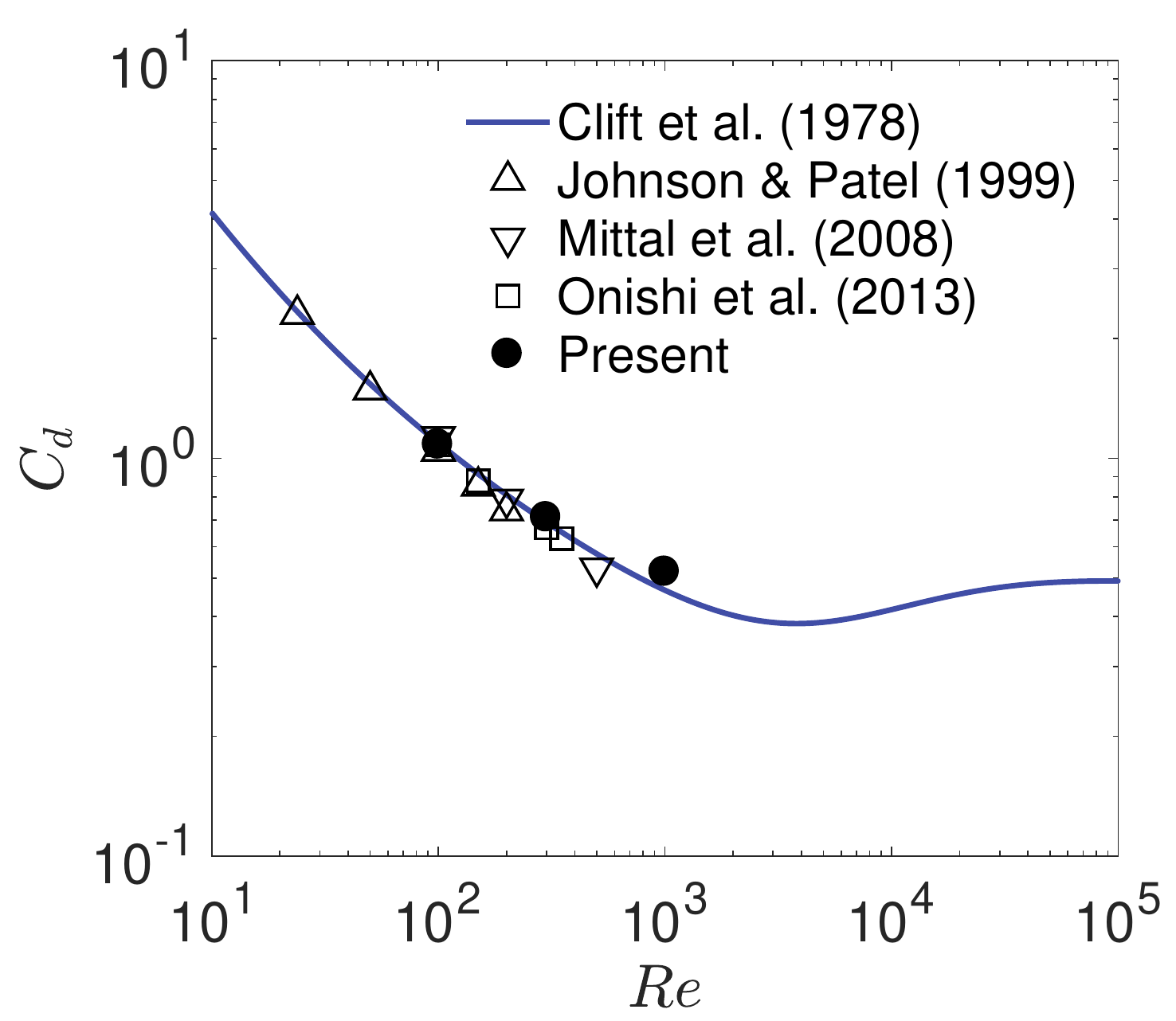}
\caption{\label{fig:Sphere-drag-coefficient} A comparison of drag coefficient of flow around a sphere with other works \cite{mitt08,onis13,john99,clif78} at different Re.} 
\end{figure}

 Following are the details of the computational domain used for the simulations: $-25D$~to~$25D$ in all three directions with the sphere placed at the center of the domain. Here, $D$ is the diameter of the sphere. The hierarchical mesh of the BCM has $l_n=7$ and $l_r=6$. The mesh spacing on the finest level $l=6$, where the sphere is placed, is $\Delta\mathbf{x}\mid_{l_{6}} = 0.012D$.  For all the simulations slip boundary condition is used on  domain boundaries in $y$ and $z$ directions, and inflow and outflow boundary condition for $x^{-}$ and $x^{+}$ boundaries, respectively. In Fig. \ref{fig:Sphere-drag-coefficient} we compare the drag coefficient with established numerical and experimental results. We find that our results are in good agreement with both numerical and experimental data. 

 Table\ref{tab:spherewake} shows a comparison of wake bubble measurements for flow at $Re=100$ with experimental and numerical data. The measurements of the wake bubble center agrees well with both numerical and experimental data, where as the bubble length shows better agreement with experiments. 
 %The symbols in the table have standard definitions. 

% \begin{figure}[th]
% \centering
% \begin{minipage}{0.88\textwidth}
% \begin{minipage}{0.49\textwidth}
% \begin{center}
% \includegraphics[width=\textwidth]{./figs/Sphere/Re100/Re100_velmag.png} {(a) Re $=100$.}
% \end{center}
% \end{minipage}
% % \hfill
% \begin{minipage}{0.49\textwidth}
% \begin{center}
% \includegraphics[width=\textwidth]{./figs/Sphere/Re300/Re300_velmag.png} {(b) Re $=300$.}
% \end{center}
% \end{minipage} 
% % \\
% % \begin{minipage}{0.49\textwidth}
% % \begin{center}
% % \includegraphics[width=\textwidth]{./figs/Sphere/Re1000/Re1000_velmag.png} {(c) Re $=1000$. }
% % \end{center}
% % \end{minipage}
% % % \hfill
% % \begin{minipage}{0.49\textwidth}
% % \begin{center}
% % \includegraphics[width=\textwidth]{./figs/Sphere/Re10000/Re10000_velmag.png} {(d) Re $=10000$. }
% % \end{center}
% % \end{minipage} 
% \end{minipage}
% % \hfill
% % \begin{minipage}{0.4\textwidth} 
% % \centering
% % \begin{minipage}{0.99\textwidth}
% % \begin{center}
% % \includegraphics[width=\textwidth]{./figs/Sphere/Sphere_Cd_Re.pdf} {(e)  }
% % \end{center}
% % \end{minipage} 
% % \begin{minipage}{\textwidth}
% % % \caption{(a)-(d) Flow around a stationary sphere at different Re plotted in velocity magnitude contours. (e) A comparison of drag coefficient of flow around a sphere with other works at different Re.}
% % \end{minipage}
% % \end{minipage}
%  \caption{\label{fig:Sphere-flow} Flow around a stationary sphere at different Re plotted in velocity magnitude contours.}
% \end{figure}

\begin{table}[h]
  \small\sf\centering
  \caption{\label{tab:spherewake} Wake bubble measurements of present work and from literature for $Re=100$.}
  \begin{tabular}{lccc}
    \toprule
    & $y_c/D$ & $x_c/D$ & $L_b/D$    \\
    \midrule
    Present &  0.288 & 0.729 & 0.794 \\ 
    \cite{mitt08} & 0.278 & 0.742 &  0.84 \\
    \cite{john99} & 0.29 & 0.75 & 0.88 \\
    \cite{tane56} & 0.28  & 0.74 & 0.8 \\ 
    \bottomrule
  \end{tabular} 
\end{table}

\subsection{Impulsively started plate}
 We validate the numerical method for a moving immersed body with the simulation of an impulsively started/accelerated infinitesimally thin plate moving perpendicular to its surface in a quiescent fluid. The experimental work of Taneda and Honji \cite{tane71} is used to validate our work. A plate of dimensions $h\times 20h$ is used in a computational domain of size $40h \times 40h \times 40h$, where $h$ is the plate height. The domain extents are: $-20h$ to $20h$ along all three directions. At $t=0$ the plate is located with its centroid at the origin of the computational domain. On the finest level of the adaptive mesh, the mesh resolution was $\Delta\mathbf{x}\mid_{l_{5}}=0.02h$. A large aspect ratio of 20 is chosen for the plate so that flow from the plate's lateral edges do not affect the flow at the center of the plate where the flow is analyzed.  The domain size is chosen such that the plate is sufficiently far from the domain boundaries  so as to avoid any boundary effects. The Plate is moved with a constant velocity $U$ in the $x^{-}$ direction. 
 
\begin{figure}[th]
 \centering
 \includegraphics[width=0.5\textwidth]{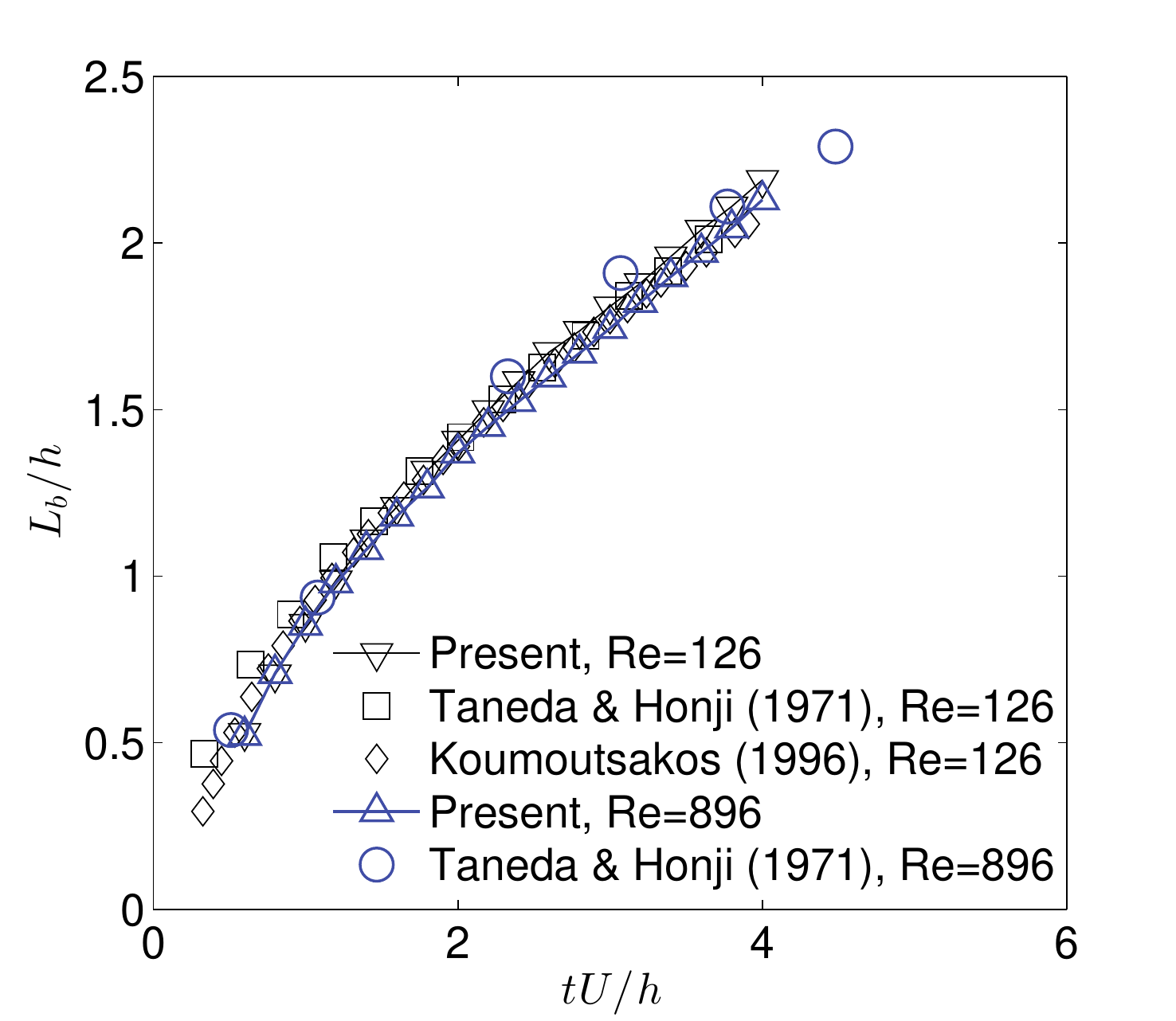}
 \caption{\label{fig:plate-wake-measurements} Evolution of wake bubble behind an impulsively stated plate at Re$=126$ and  Re$=896$. The simulations show good agreement with Kaoumatsakos' \cite{koum96} 2D simulation at $Re=126$ and,  Taneda \& Honji's \cite{tane71} experimental results.}
 \end{figure}

 The simulation was carried out at $Re=126$ and $Re=896$, where plate height and velocity are used as characteristic length and velocity, respectively, to define the Reynolds number. A comparison of wake bubble size from the simulations with  Taneda and Honji's \cite{tane71} experimental data is shown in Fig. \ref{fig:plate-wake-measurements}. The figure shows evolution of wake bubble size ($L_b$) normalized by plate height $h$ as a function of dimensionless time $tU/h$.  We also compare our results with the $2D$ simulation results of Koumoutsakos' \cite{koum96} at $Re=126$. Our results are in good agreement with both experimental and simulation data from literature. 
 
\section{Software Environment}
\label{sec:framework}
Based on the Finite Volume Method, we have developed a unified solver framework \textsc{Cube} (Complex Unified Building cubE) for solving large-scale compressible and incompressible flow problems. The framework has a modular design where \textsc{Cube} provides a core library containing kernel functionalities e.g. mesh representation, numerical kernels for computing flow field and I/O routines. Solvers are then developed on top of the kernel by connecting necessary kernel modules together, forming a ``solver'' pipeline, describing the necessary steps to solve a particular problem.

\subsection{Object-Oriented abstraction}
Written in modern Fortran 2003, \textsc{Cube} employs a lightweight Object-Oriented abstraction. A set of abstract classes/types define canonical components of a solver, which are later provided by a real solver. With a set of predefined solvers and parameters, \textsc{Cube} can be used as a regular CFD solver. Meanwhile, for advanced users it is also possible to develop their own tailor-made applications by overloading certain kernel components similar to other C++ based solver frameworks, e.g. FEniCS \cite{fenics15} and OpenFOAM \cite{foam}.

%%  Since the framework is targeted to different kinds of users, \textsc{Cube}'s framework already comes with a set of predefined solvers, intended for a general user. By only providing simulation parameters and geometries, \textsc{Cube} can be used as a regular flow solver.

%% Advanced users can instead see \textsc{Cube} as a library. And use it to develop own solvers, tailor-made for a particular application. Similar to other C++ based solver frameworks, e.g FEniCS \cite{fenics15} and OpenFOAM \cite{foam}, \textsc{Cube} let advanced users overload/implemented certain kernel components in their application code.

For example, if a user wishes to write her own application specific flux evaluation routine, she would only need to overload the base definition of a numerical flux \verb|flux_t| (Fig. \ref{fig:absfluxes}) inside the kernel. Once done, the framework will execute the user's code (in this case the \verb|eval| subroutine) instead of the one in the kernel every time fluxes are evaluated. This way we can keep \textsc{Cube}'s kernel small and general, without application specific code.
\begin{figure}[h]
\centering
 \begin{tcolorbox}[colframe=rikenred,boxsep=-5pt]
\begin{lstlisting}[frame=none,language=Fortran,basicstyle=\fontsize{7}{7}\selectfont\ttfamily]
type flux_t
   real(kind=dp), allocatable :: f(:,:,:,:,:)
   procedure(scheme), nopass, pointer :: eval
end type flux_t
\end{lstlisting}
%\lstinputlisting[frame=none,language=Fortran,basicstyle=\footnotesize]{data/fluxes.f90}
  \end{tcolorbox}
\caption{\label{fig:absfluxes} An illustration of the abstract definition of a numerical flux inside \textsc{Cube}.}
\end{figure}

%% The coordinate of the cell center defined by the index $\mathbf{i}=(i,j,k)$, in the $t^{th}$ cube, is give by
%% \begin{equation}
%%  \mathbf{x_i}^{t} = \mathbf{x}_{c}^{t} + \left( (i+\frac{1}{2})\Delta x_{1}^{t}, (j+\frac{1}{2})\Delta  x_{2}^{t}, (k+\frac{1}{2})\Delta  x_{3}^{t} \right),
%%  \label{eqn-cell-center}
%% \end{equation}
%% where $\mathbf{x}_{c}^{t} = (x_{1c},x_{2c},x_{3c})$ is the coordinate of the base corner of a cube indexed $t$, and $\Delta\mathbf{x}^{t} = (\Delta x_{1}^{t}, \Delta x_{2}^{t}, \Delta x_{3}^{t})$ is the mesh spacing. Base corner of a cube is the corner with the minimum coordinate along the principle directions. $\mathbf{x}_{c}$ is the set of base corner coordinates of all the cubes in a mesh so that $\mathbf{x}_{c}^{t} \in \mathbf{x}_{c} $. Similarly, $\Delta\mathbf{x}^{t} \in \Delta\mathbf{x}$, where $\Delta\mathbf{x}$ is the set of mesh spacing of each cube in the BCM mesh.

\section{Enabling Large Scale Simulations}
\label{sec:para}
\subsection{Data Decomposition}
\textsc{Cube} has been parallelized using a hybrid MPI + OpenMP approach, where whole cubes are subdivided between MPI ranks, and thread parallelization of the numerical kernels are performed on a per cube basis or with two dimensional slices in the z-direction of each cube. MPI partitioning is performed following a load-balanced linear data distribution \cite{Vel1994a}; Let $P$ be the number of MPI ranks, p be an MPI rank, $N$ be the global number of cubes, and $n$ be the local number of cubes assigned to a rank. With $N = P L + R$ and $0 \leq R < P$, we have that
\begin{equation}
\label{eq:loaddist}
\begin{split}
  L  = &\left \lfloor \frac{N}{P} \right \rfloor, \\
  R  = & \,\, N \bmod P, \\
  n = & \left \lfloor \frac{N + P - p - 1}{P} \right \rfloor.
\end{split}
%  \mu(m) & = (p,i), \, \textrm{where} \left\{
%  \begin{array}{ll}
%    p = \max\left(  \left \lfloor \frac{m}{L+1} \right \rfloor ,   \left \lfloor \frac{m - R}{L} \right \rfloor \righ\t) & \\
%    i = m - pL - \min(p,R) &
%  \end{array}
%\right.
\end{equation}
To minimize data dependencies, the distribution is calculated based on a space-filling curve, the Z-ordering (cf. \cite{Bader2013}) of the cubes given by our mesh generator. Using the linear distribution as defined in \eqref{eq:loaddist}, we can easily calculate an owner of a cube given its global id. Therefore, each MPI ranks builds a look-up table for local to global index mappings, which can later be used to construct adjacency information on the fly. If the Z-ordering is not sufficient, a new distribution can be calculated using ParMETIS \cite{SchKar1997a} or the built in load balancing framework (see Section \ref{sec:loadb}).

\subsection{Lagrangian domain decomposition}
\begin{figure}[th]
  \centering
  \null\hfill
   \subfloat[Unpartitioned mesh.]{\label{fig:EL}\includegraphics[width=0.45\columnwidth]{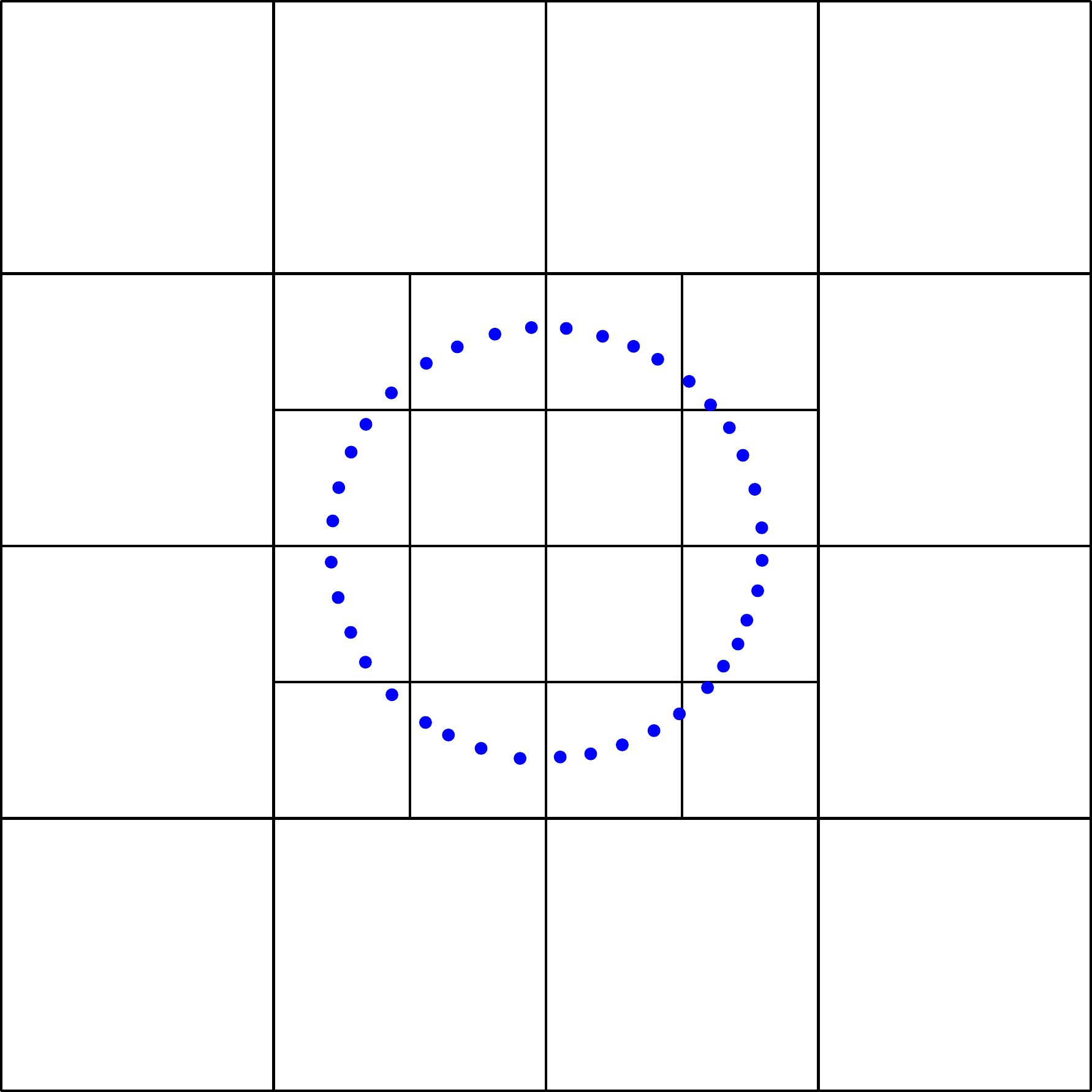}}
   \hfill
   \subfloat[Partitioned mesh.]{\label{fig:EL_decomp}\includegraphics[width=0.45\columnwidth]{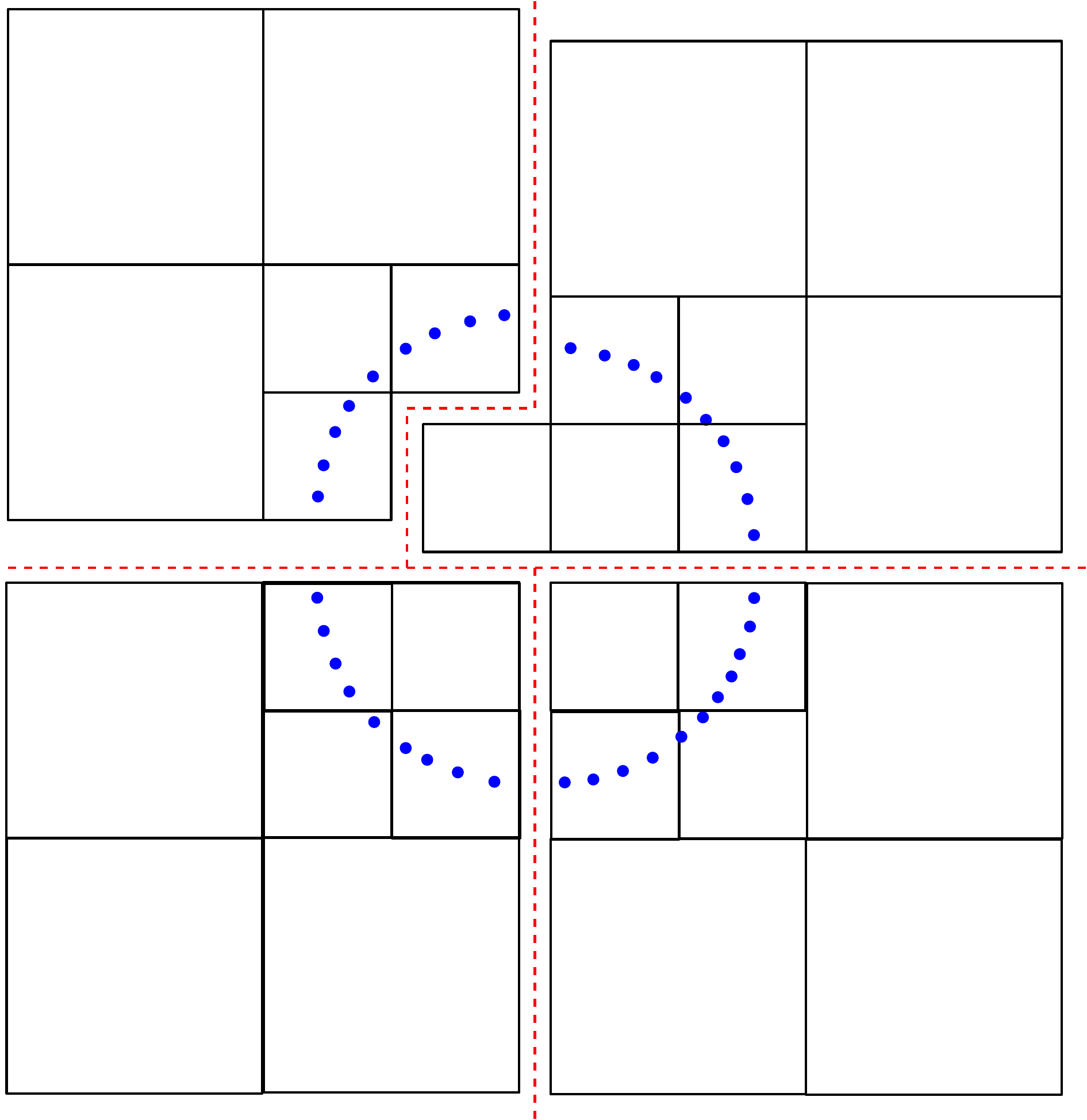}}
   \hfill\null
  \caption{\label{fig:EL_dec_str} An unpartitioned BCM mesh with lagrangian particles \protect\subref{fig:EL}. BCM mesh and the Lagrangian particles partitioned into 4 paritions \protect\subref{fig:EL_decomp}. Separation of the paritions is marked by the dashed red lines. }
\end{figure}

For a Lagrangian-Eulerian framework, such as the one employed in the present work for modeling immersed geometries, there are more than one methods of decomposing the system. As there are two separate domains, namely the Lagragian domain and Eulerian domain, combined or separate domain decomposition strategies may be employed. Three of the most commonly used approaches for decomposing a Lagrangian-Eulerian system are task parallelism, atomic decomposition, and spatial decomposition \cite{kuan13}. In task parallelism the Lagrangian and Eulerian workload are assigned to separate set of processors. In atomic decomposition, after the Eulerian domain is decomposed, the Lagrangian domain is divided equally among all the processors. Spatial decomposition decomposes the Lagrangian domain on the basis of the Eulerian decomposition. Both task parallelism and atomic decomposition, due to non locality of the Eulerian and Lagrangian domain, necessitate communication between processors for Lagrangian-Eulerian interaction \cite{give06}. This is not the case with spatial decomposition. In the present work we employ the spatial decomposition.  A schematic of spatial decomposition applied to combined system of the Lagrangian particles and the cubes of the BCM mesh is shown in Fig. \ref{fig:EL_dec_str}. In the schematic the partitioning of the BCM mesh is highlighted by the dashed red lines between the cubes, as shown in Fig. \ref{fig:EL_decomp}. As described in Sec. \ref{sec:bcl}, Lagrangian particles are grouped into a $set$ for each cube of the BCM mesh. Lagrangian particle $set$s of cubes belonging to partition $p$  are assigned to the same partition $p$.

% \begin{comment}
\begin{algorithm2e}[t]
  \SetNoFillComment
  \SetAlgoNoLine
  \SetKwProg{Fn}{Function}{:}{}
  \Fn{Exchange}{
    \eIf{first thread to arrive}{
      Post Irecv\;

      \For{each cube in the halo}{
        Pack data\;
      }
      Post Isend\;
    }{
      \For{each cube}{
        Local halo exchange\;
      }
    }
  }

  \Fn{Finalize}{
    \While{Halo data has not arrived}{
      Check if data has arrived (Testsome)\;

      \For{each received halo}{
        Add halo data to local cubes\;
      }
    }
}
\caption{\label{alg:halo} Multithreaded halo-exchange}
\end{algorithm2e}
% \end{comment}

\subsection{Halo-exchange}
A key to achieve good performance in block structured codes is an efficient halo-exchange routine, locally (between cubes owned by a MPI rank) as well as globally between adjacent MPI ranks. There are many factors that can impact performance, such as the width of the halo or the number of cells per cube \cite{Olschanowsky2014}. Increasing the number of cells per cube is a less viable approach for \textsc{Cube}. Since we target industrial applications, complex geometries will require very fine resolution in very localized areas of the domain. For a given mesh resolution around a geometry, cubes with greater number of cells would greatly increase the  overall mesh size, and quickly exhaust memory resources. Therefore, the most feasible option for our applications is to develop a highly concurrent halo-exchange algorithm, that can handle situations where large amount of halo data has to be exchanged efficiently.

To mitigate this problem, we have developed a highly concurrent multithreaded exchange algorithm, that will overlap packing, sending and unpacking of data with local halo-exchange and other possible work (if the numerical scheme permits). The pseudo-code of the algorithm is given in Algorithm \ref{alg:halo}, and is divided into two separate functions. When \textsc{Cube} needs to exchange data, the function \textit{Exchange} is called. The first thread to arrive starts filling communication buffers and posting non-blocking send/recv requests. While the first thread is busy packing and sending data to MPI neighbors, all other threads will perform the exchange of data between cubes local to an MPI rank.
%\rbcom{Would be good to show that this approach is faster than the alternate approach with raw data. Do we have data to shown that on tread packing while other do local halo exchange it faster than all threads pack and all threads do local halo exchange in sequence.}

The second function \textit{Finalize} waits until halo data has arrived, and adds contribution from received data to the local cubes. To increase the concurrency, the routine will process messages on a first come first served basis, by issuing \verb|MPI_Testsome| calls to pick the first finished recv requests from the message queue.

% Using this exchange routine, it is possible to achieve good strong scalability with \textsc{Cube}, down to few BCM cubes per MPI rank, as illustrated in Figure \ref{fig:speedup} where the result of a strong scaling test of \textsc{Cube}'s compressible solver on the K computer is shown. The result shows that good scaling is achieved down to less than two cubes per core, despite the very wide halos (four cells) required for compressible applications.
% \begin{figure}[th]
% \centering
%   \input{data/speedup}
% \caption{\label{fig:speedup} Strong scaling results on the K computer for the compressible solver in \textsc{Cube}, solving a problem consisting of 56142 cubes with a four cell wide halo and 32 cells per cube.}
% \end{figure}

\subsubsection{Overlapped time-stepping}
The halo-exchange wait time between \textit{Exchange} and \textit{Finalize} until the data has arrived, creates a serial section and severely limits the potential scalability. This overlap window could be utilized to perform operations that do not require information from other MPI ranks. The easiest option to utilize the time between non-blocking send and recive calls is to impose boundary conditions, which does not require any communcation. However, typically, imposition of boundary conditions is not expensive, and will not cover the entire overlap window.

\begin{figure}[th]
\centering
\subfloat[Subdivision of the grid into internal (blue) and external (red)  zones for four MPI ranks.]{\label{fig:zones}
 \includegraphics[width=0.6\columnwidth]{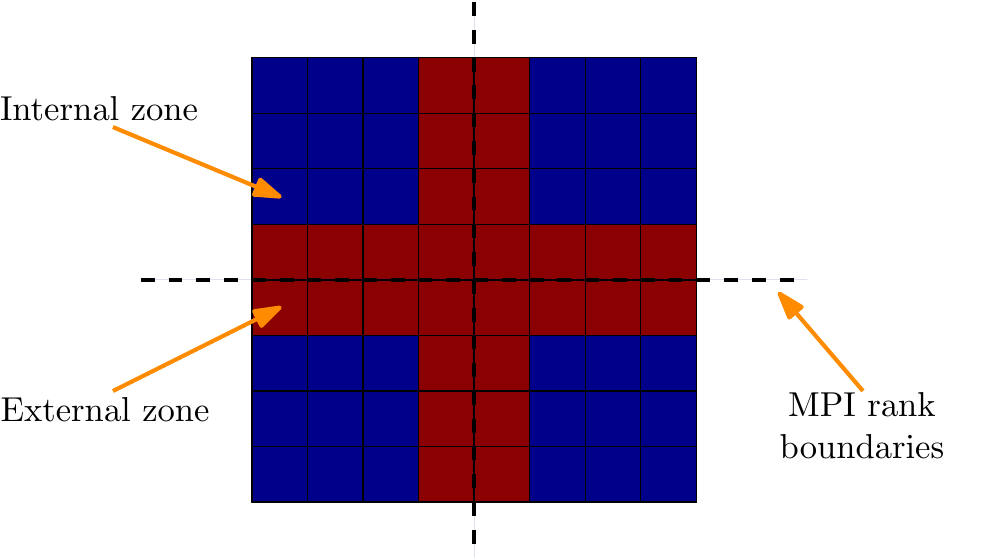}}
% \hspace{1em}
\subfloat[Time per time-step.]{\label{fig:overlapped}
 \includegraphics[width=0.25\columnwidth]{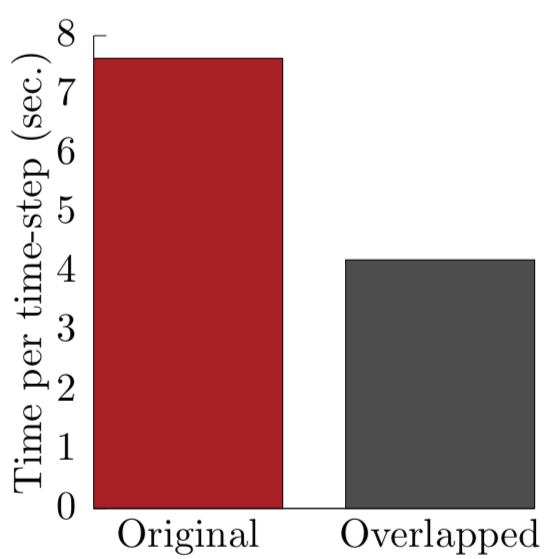}}
  \caption{Time per time-step \ref{fig:overlapped}, comparing ordinary time-stepping with overlapped for simulation of flow past a full car on the K computer, running on 128 MPI ranks using 8 threads each.}
\end{figure}

In order to enable utilization of the overlap window in Alg. \ref{alg:halo}, we subdivide the local cubes on an MPI rank into two different zones, namely \textit{internal zone} and \textit{external zone}. \textit{Internal zone} is a spatial region containing only those cubes all of whose neighbours are on the same rank. And, we define \textit{external zone} as a region containing cubes with at least one off rank neighbour as shown in Fig. \ref{fig:zones}. With such a zoning of cubes, we can extend the overlapped halo-exchange to all stages of the solution algorithm, filling the wait time window between \textit{Exchange} and \textit{Finalize} with as much work as the numerical method permits.

%The wait time between \textit{Exchange} and \textit{Finalize} until the data is received in \textit{Finalize} is wasted time, so it could be utilized to perform operations that do not require MPI neighbor halo information. The easiest option to utilize the wait time between non-blocking send and recieve posts is to impose the domain boundary conditions which does not require MPI neighbor halo information. But, typically, imposition of domain boundary condition is not computationally intensive, so the wait time may not be fully utilized fully. Filling this wait time gap between \textit{Exchange} and \textit{Finalize} with work that may comsume more time than the available time gap is reasonable so long as the wait time gap is fully utilized to perform useful work.

%In order to enable utilization of the wait time (or time gap) window between \textit{Exchange} and \textit{Finalize} in Algorithm \ref{alg:halo}, we subdivide the local cubes on an MPI rank into two different zones, namely \textit{internal zone} and \textit{external zone}. \textit{Internal zone} is a spatial region containing only those cubes all of whose neighbours are on the same rank. And, we define \textit{external zone} as a region containing cubes with at least one off rank neighbour as shown in Fig. \ref{fig:zones}. With such a zoning of cubes, we can extend the overlapped halo-exchange to all stages of the solution algorithm, filling the wait time window between \textit{Exchange} and \textit{Finalize} with as much work as the numerical method permits.

Here we describe an overlapped time-stepping method that can make full utilization of the overlap window between halo exchange routines. In the numerical algorithm of the present work (Sec. \ref{sec:n-alg}), the halo exchange needs to be performed during first substep (Eq. \ref{eq:substep1}) and during the iterative solution of the Poisson equation. During these two steps we can overlap halo exchange with core computation of the solver. The overlapped time-stepping algorithm is shown in Alg. \ref{alg:overlapt}. At each time-step, in the function \textit{FracTimeStep} the solver first initiates the halo-exchange by calling \textit{Exchange}. Once all threads have returned from \textit{Exchange}, there will exist enough valid data locally to compute Eq. \ref{eq:substep1} in all cubes belonging to the internal zone. After the internal zone calculation is complete, the \textit{Finalize} function is called to distribute data that has already been received to the relevant external zone cubes. Once the rank neighbor halo information has arrived, the same calculation is performed in the external zone thereby completing substep one (Eq. \ref{eq:substep1}) of the numerical algorithm. Here, an explicit Adams-Bashforth time-stepping was assumed in the description of overlapped time-stepping of the \textit{FracTimeStep}. Following the completion of the \textit{FracTimeStep}, the solver can continue to the next substep of the numerical algorithm, the Poisson solver. In the \textit{PoissonSolve} function, pressure is solved iteratively until the error in pressure converges to a specified tolerance. Within each of these iteration we perform the overlapped halo exchange as described above for \textit{FracTimeStep} (Algorithm \ref{alg:overlapt}).

\begin{algorithm2e}[t]
  \SetNoFillComment
  \SetAlgoNoLine
    \SetKwProg{Fn}{Function}{:}{}
  \While{(time $<$ time end)}{
    \textbf{call} \textit{FracTimeStep} \;        

    \textbf{call} \textit{PoissonSolve} \;        

    \textbf{call} \textit{Velocity Correction} \;        
  }

  \BlankLine
  \Fn{FracTimeStep}{
    \textbf{call} \textit{Exchange}(Velocity) \;
    
    \Indp\textbf{call} \textit{fluxes}(internal$\_$cubes) \;
    
    \textbf{call} \textit{Integrate$\_$in$\_$time}(internal$\_$cubes) \;
    
    \Indm \textbf{call} \textit{Finalize}(Velocity) \;

    \Indp\textbf{call} \textit{fluxes}(external$\_$cubes) \;
    
    \textbf{call} \textit{Integrate$\_$in$\_$time}(external$\_$cubes) \;
    }
  \BlankLine
    \Fn{PoissonSolve}{
      \While{Pressure not converged}{
        \textbf{call} \textit{Exchange}(Pressure) \;

        \Indp \textbf{call} \textit{Update Pressure}(internal$\_$cubes) \;

        \Indm \textbf{call} \textit{Finalize}(Pressure) \;

        \Indp \textbf{call} \textit{Update Pressure}(external$\_$cubes) \;

      }
    }

\caption{\label{alg:overlapt} Overlapped time-stepping}
\end{algorithm2e}

%% \rbcom{Need a more detailed description.\\
%%        Do we have more data points for this? \\
%%        Show the improvement at different mpi ranks?\\
%%        Would be good to add an algorithm describing the overlapped time stepping.}

The potential of utilizing the full overlapped time-stepping scheme is illustrated in Fig. \ref{fig:overlapped}, where the time to solution for an incompressible simulation of a full car model (see Fig. \ref{fig:cervo}) is reduced by almost a factor of half when running on 128 nodes on the K computer with a mesh of 47811 cubes. %\rbcom{Some details about the mesh size, what kind of simulation, what geometry was uased could be added here.}

\subsection{I/O strategies}
\label{sec:iostart}
Efficient I/O strategies have in recent years become a very important component of a scalable code. Not only should the I/O routines be able to dump data as fast as possible, but loading input data and  restarting from checkpoint files are as important. For large scale simulations, the ability to load/restart from previous runs on any number of MPI ranks without offline processing is also a necessity.

To accommodate these requirements \textsc{Cube} implements parallel I/O in form of MPI I/O. Any kind of data is written to a shared file as a flat binary stream. A small header in the beginning of each file contains necessary information to compute new load-balanced distributions using equation \eqref{eq:loaddist}, which allows for the data to be read back on any number of ranks automatically.

\subsubsection{Data compression}
The high spatial resolution required for accurate simulations can easily exceed hundreds of millions of grid points. For such problems, the data size quickly becomes a problem. While quickly writing data to disk is not the biggest concern, the problem is the psychical disk space, which can easily be saturated for a time dependent problem.

To solve this problem, we use the discrete wavelet based lossy data compression algorithm developed by Sakai and co-workers \cite{Sakai2010}. In this algorithm, a discrete wavelet transform is computed independently for each cube, which we extend to include the halo region. One level of quantization is applied and the entire signal is encoded using Zlib \cite{zlib}. The result is a stream of compressed data for each cube of various length. All these streams are combined into one large stream, which together with a header data is written to disk as a single file. Since we record the size of each cube's compressed byte stream it is straightforward to uncompress the data on any number of cores when loading/restarting from already written data. Also, since we use a hybrid parallelization, the compression algorithm is also multithreaded such that each thread will transform and encode independent cubes, and a final reduction will combine all temporary streams from the threads into the stream that will be written to disk. In total, all necessary operations to compress our data account for about one forth of the total I/O time for writing data to disk.

 \begin{algorithm2e}
   \SetNoFillComment
   \SetAlgoNoLine
   \For{$i \hdots n_{cube}$}{
     $j \leftarrow i + (thrd_{id} - 1)$\;

     \If{$j <= n_{cube}$}{
       dwt$_{\textrm{stream}}$ $\leftarrow$ dwt(cube$(j) $)\;

       quant$_{\textrm{stream}}$ $\leftarrow$ quantification(dwt$_{\textrm{stream}}$)\;

       encoded$_{\textrm{stream}}$ $\leftarrow$ encode(quant$_{\textrm{stream}}$)\;
     }
     $i \leftarrow i + n_{thrds}$\;
  }
\caption{\label{alg:comp} Compression algorithm}
\end{algorithm2e}

\begin{figure}[th]
\centering
    \subfloat[Compression ratio test.]{\label{fig:iocomp}\includegraphics[width=0.32\columnwidth]{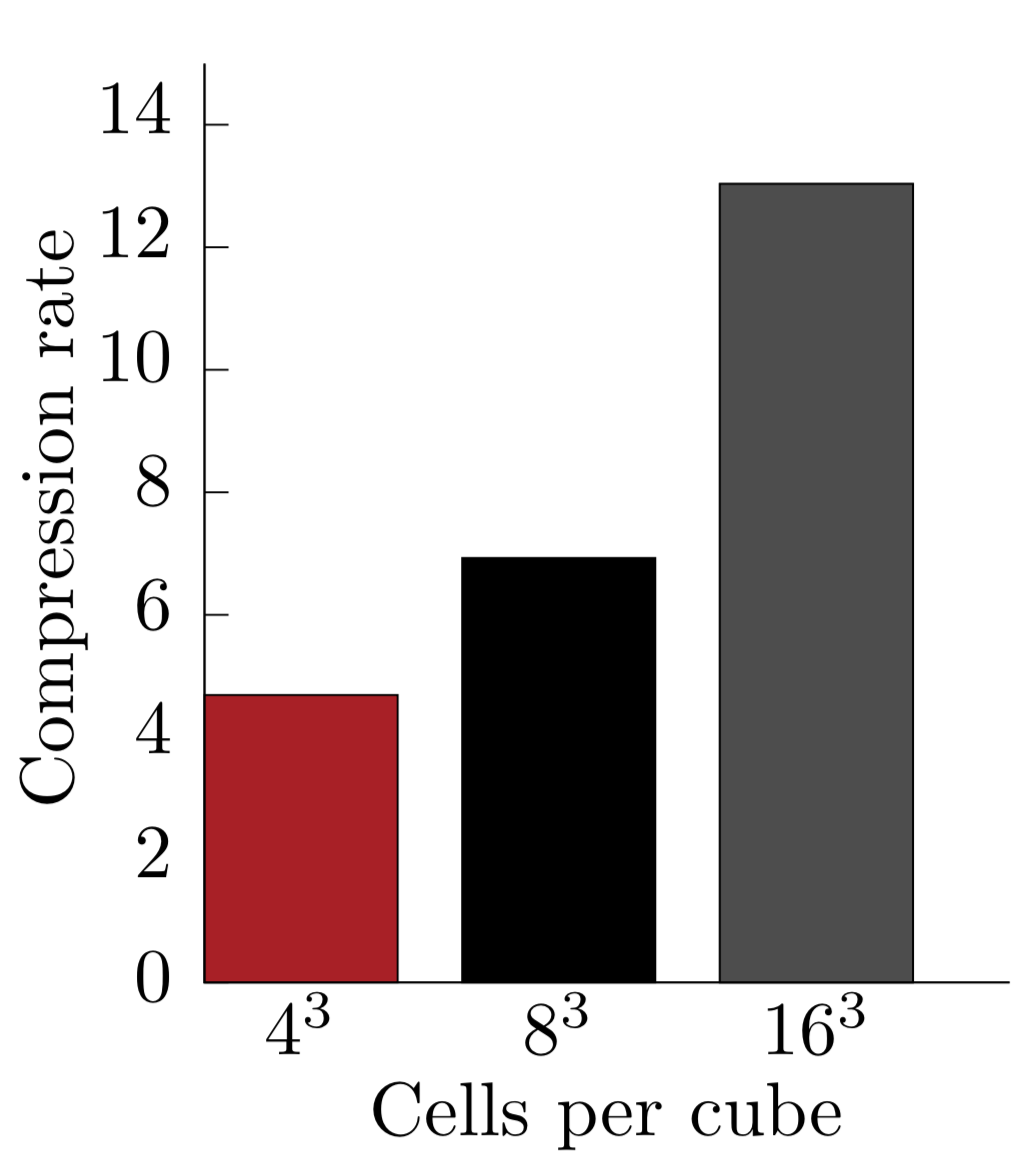}}
    \hspace{0.1\textwidth}
    \subfloat[I/O thorughput.]{\label{fig:iobw}\includegraphics[width=0.32\columnwidth]{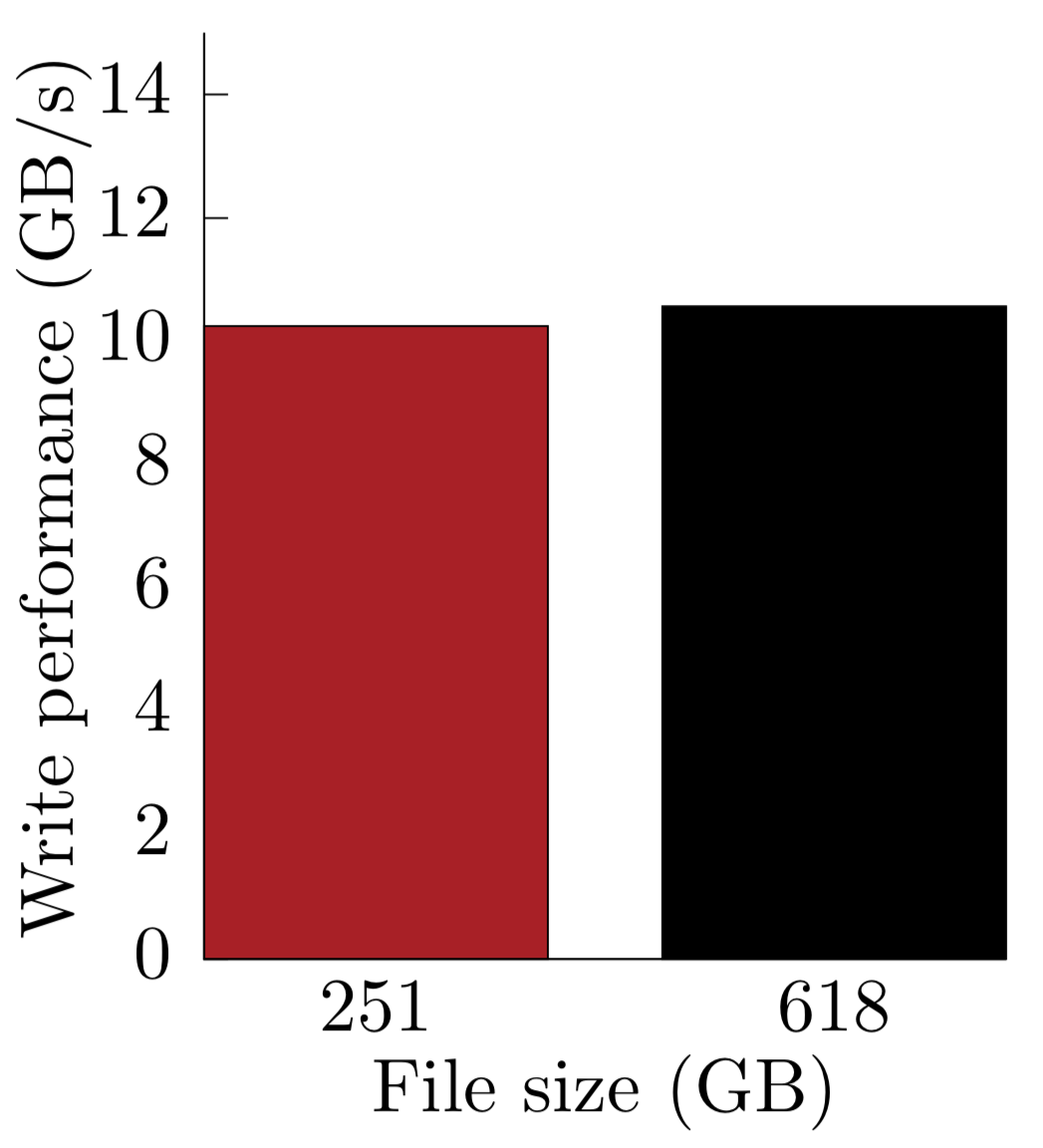}}
  \caption{Achieved compression ratio for various cell discretizations \protect\subref{fig:iocomp} and I/O throughput \protect\subref{fig:iobw} for a typical application running on 16385 MPI ranks on the K computer, for two different cube discretization of a mesh consisting of 32M cubes.}
\end{figure}

The achieved compression ratio depends on at least two factors, the number of cells per cube and information loss introduced by the quantization. For an error tolerance of $O(10^{-4})$, which is sufficient for a lossy checkpoint restart, the compression ratio ranges between $\approx 1$:$4 - 1$:$15$ for $4^3 - 16^3$ cells per cube (see Fig. \ref{fig:iocomp}), and for higher errors $O(10^{-2})$ a ratio of up to $1:43$ can be achieved. %\rbcom{do we have data, which we can plot, for compression ratio when error are $O(10^{-2})$ or higher? } No we don't have any good data

For a typical application, the I/O throughput is on an average around 10GB/s on the K computer (see Fig \ref{fig:iobw}), which is far away from the K's theoretical peak bandwidth, but well above the average throughput of most HPC applications \cite{Luu2015}.

However, if data compression is not sufficient, \textsc{Cube} also includes in-situ visualization through VisIt \cite{Childs:2005:ACS}, either in traditional interactive mode or in batch mode, where  a user can setup a user-defined pipeline that will be rendered during the simulation directly from the simulations memory without any data copying or I/O operations.

\subsection{Load Balancing}
\label{sec:loadb}
Load balancing is an essential component in today's large scale multiphysics simulations, and with an ever increasing amount of parallelism in modern computer architecture it is essential to reduce even the slightest workload imbalance. An
imbalance could severely impact an application's scalability. Traditionally, load balancing is seen as a static problem, closely related to the fundamental problem of parallel computing, namely data decomposition. For a CFD simulation based on BCM, since each cubes contains the same amount of cells the goal is to evenly distribute the cubes among the available cores. However, such a decomposition assumes that the workload for each cube is uniform. For most cubes this is true, but for cubes which contain Lagrangian particles the workload is slightly higher, which implies a workload imbalance. Therefore,  to retain good scalability we need to derive a load balancing method  that balances the workload not only considering the Eulerian computational mesh, but also the additional workload from the immersed body.

\subsubsection{Static Load Balancing}
\label{sec:data}
In parallel computing, the idea of data decomposition or static load balancing is simple, namely divide the workload evenly across all the workers. This can be formulated as the partitioning problem.

Given a set of cells $\mathcal{C}$ from a domain $\mathcal{T}$, the partitioning problem for $p$ workers can be expressed as, find $p$ subsets  $\lbrace \mathcal{T}^i \rbrace_{i=1}^p$  such that:
\begin{equation}
  \label{eq:partprob}
  \mathcal{T} = \cup_{i=1}^p \mathcal{T}^i, \quad \textrm{and} \quad \mathcal{T}^i \cap \mathcal{T}^j = \emptyset\, \quad \textrm{when }   i \neq j,
\end{equation}
with the constraint that the workload:
\begin{displaymath}
  W(\mathcal{T}^i) = |\{\mathcal{C} \in \mathcal{T} \mid \mathcal{C} \in \mathcal{T}^i\}|,
\end{displaymath}
should be approximately equal for all subsets.

Solving Eq. \eqref{eq:partprob} can be done in several ways. The least expensive geometric methods, such as space filling curves \cite{Bader2013}, only depend on the geometry of the domain. These methods are fast, but don't take into account the topology, hence there is a data dependencies between different cells in the domain. For Cartesian meshes such as BCM, neglecting to consider data dependencies will not lead to severe imbalances. In the scenario where all the cells have the same amount of neighbors, if the decomposition method tries to assign cells, which are close to each other, to one worker (in the geometrical sense) then the data dependencies will ``automatically'' be approximately balanced. However, if cells have a non uniform workload, or the problem has asymmetric data dependencies between cells, we have to resort to graph methods in order to solve Eq. \eqref{eq:partprob}.

Graph methods don't solve Eq. \eqref{eq:partprob} directly, instead they consider the $k$-way partitioning problem. To understand $k$-way partitioning, consider an undirected graph $G = (V, E)$, with nodes $V$ and edges $E$. The nodes $V$ are split into $k$ subsets $\lbrace Q_j \rbrace^k_{j=1}$ with the constraint that the number of nodes should be roughly equal in each subset, and the number of edges cut should be minimized. If we model the computational work by $V$ and the data dependencies in the domain by $E$, we see that this method will balance both the computational work and the dependencies. Furthermore, if we instead consider a weighted graph $G$ and add the constraint that the sum of all weights should be roughly equal in all subsets $Q_j$, the method can then, by allowing different weights in the graph, handle a non uniform workload.

\subsubsection{Dynamic Load Balancing}
\label{sec:dyn}
In order to perform dynamic load balancing two components are needed. First, a way to evaluate the workload and second, a way to decompose the data with the constraint to even out the workload. Using graph based methods from Section \ref{sec:data} we can compute new constrained partitions of our computational domain. But the challenge is to be able to evaluate the current and future workloads, and decide if load balancing is needed.

\begin{figure}[th]
%  \begin{center}
%  \subfloat[Building Cube mesh.]{\label{fig:mesh}\includegraphics[width=0.4\columnwidth]{mesh}}
%  \hspace*{0.1\textwidth}
%  \subfloat[The dual graph.]{\label{fig:dual}\includegraphics[width=0.4\columnwidth]{graph_dual}}
%  \end{center}
  \centering
  \includegraphics[width=0.8\columnwidth]{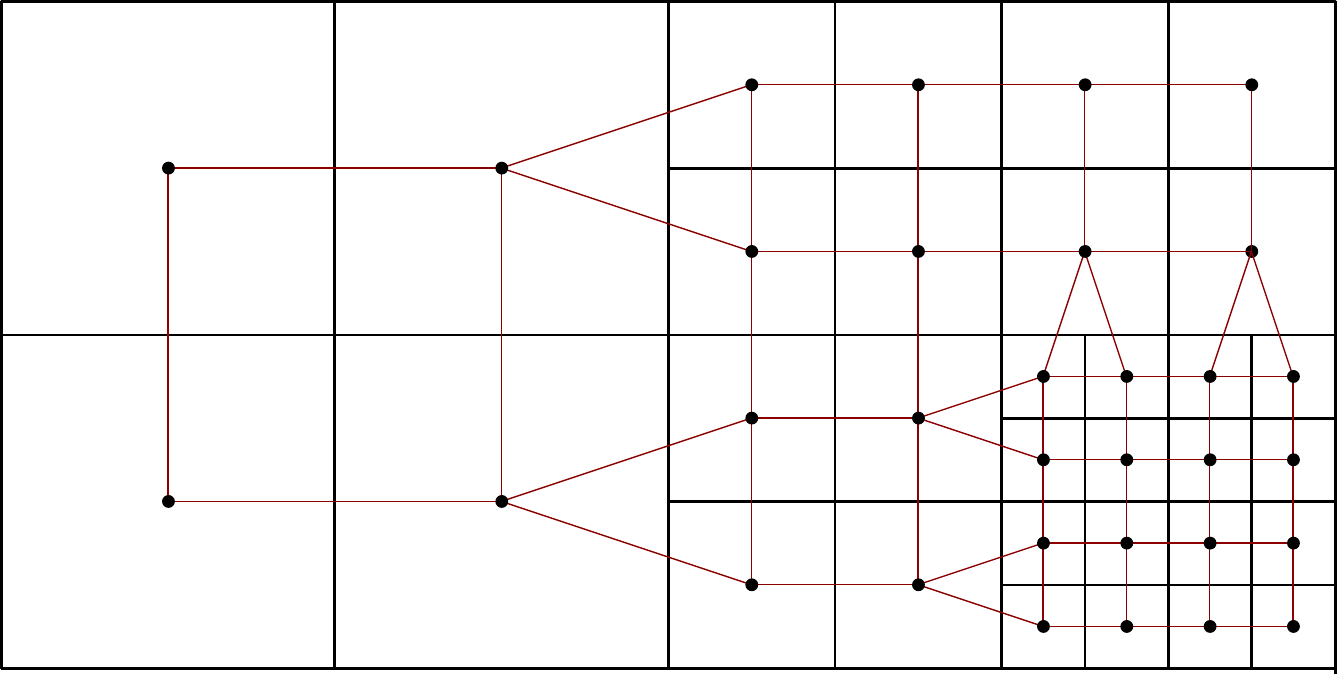}
  \caption{\label{fig:dualgraph} Example of the dual graph of a Building Cube mesh.}
\end{figure}

% \begin{comment}

% \end{comment}

\subsubsection{Workload Modeling}
\label{sec:work}
We model the workload by a weighted dual graph of the underlying Building Cube mesh (see Fig. \ref{fig:dualgraph}). Let $G = (V, E)$ be the dual graph of the mesh, with nodes $V$ (one for each cube) and edges $E$ (connecting two nodes if their respective cubes share a common face), $q$ be one of the partitions and let $w_i$ be the computational work (weights) assigned to the graph. The workload of a partition $q\in\mathcal{T}$ is then defined as:
\begin{equation*}
\label{eq:workload}
W(q) = \sum_{w_i \in w_q} w_i
\end{equation*}
 Let $W_{\textrm{avg}}$ be the average  workload and $W_{\textrm{max}}$ be the maximum, then the graph is considered unbalanced if:
 \begin{equation*}
\label{eq:imb}
   W_{\textrm{max}} / W_{\textrm{avg}} > \kappa
 \end{equation*}
where $\kappa$ is the threshold value, determined depending on the problem at hand and/or machine characteristics.

To model a simulation's workload we finally have to assign appropriate values to the graph's weights $w_i$. In order to have a fine grained control over the workload, we let each node have $j$ weights $w^{v_j}_i$, representing the computational work for the given node, and we let  each edge have $k$ weights $w^{e_k}_i$ representing communication costs (data dependencies between graph nodes). The total weight for a given graph node is then given by,
\begin{equation}
  \label{eq:nodweight}
  w_i = \sum_j w^{v_j}_i + \sum_k w^{e_k}_i.
\end{equation}

For a typical simulation, we assign the number of grid points in each cube to $w^{v_1}_i$ and the size of the halo (number of grid points to exchange between cubes) to $w^{e_k}_i$ for each of the graph edges connected to node $V_i$. Additional weights can also be added to the edges, but in the present study we limit ourselves to modeling only the halo exchange cost. We add an additional weight to the graph's node to model the additional computational cost of the immersed bodies. The additional immersed body cost $w_i^{v_{2}}$ is added in Eq. \eqref{eq:nodweight}, which is modeled as a $\gamma \cdot n_\textrm{particles}$. Here, $\gamma$ is a cost parameter for the particles and $n_{particles}$ is the number of particles in each cube.

%\subsubsection{IB Workload Parameter}
The operations related to the immersed bodies involve computationally intensive operations of interpolation and projection between the Eulerian and the Lagrangian meshes, and is done once every time-step. Unlike $w^{v_1}_i$, which can be modeled as the number of cells in a cube, $w^{v_2}_i$ cannot be modeled directly by the number of particles in cube. This is because  interpolation and projection of information involves operations in a small box of Eulerian cells ($n\times n\times n$, where $n$ depends on the discrete delta function) for each Lagragian particle. $\gamma = n^{3}$ would be an appropriate choice if interpolation and projection operations were carried out inside the pressure solver's iterations and in the Crank-Nicholson scheme's iterative loop. But, the interpolation is carried out once every time-step outside the pressure solver and the temporal integrator. Consequently, it is not clear what value of $\gamma$ would be optimal. We choose $\gamma$ in the range of $1-4$ to investigate an optimal value through a parametric study in sections to follow.

\begin{figure}
 \begin{center}
   \subfloat[Unbalanced (Z-ordering).]{\label{fig:unbalanced}\includegraphics[width=0.49\columnwidth]{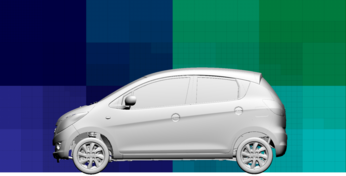}}
   \hspace*{1mm}
   \subfloat[Balanced wrt. geometry.]{\label{fig:balanced}\includegraphics[width=0.49\columnwidth]{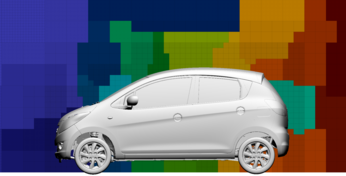}}
 \end{center}
  \caption{\label{fig:cervo-loadb} Load balancing wrt. immersed geometry and fluid cells, colored by MPI rank.}
\end{figure}

The graph is finally partitioned by a graph partitioner, with the weights as an additional balancing constraint. Once new partitions have been obtained intelligent remapping is used to assign new partitions in such a way that data movement is kept at a minimum. This is achieved by solving the maximally weighted bipartite graph problem (MWBG). A bipartite graph is one in which edges are weighted by the amount of data to be transferred with the old partitions to the right and new partitions to the left. The MWBG problem is solved using a heuristic with a linear runtime \cite{JanHof2010c,Jansson2013a}. The new load balanced partitions are obtained as illustrated in Fig. \ref{fig:cervo-loadb}. The entire load balancing framework is given in Alg. \ref{alg:dyn}.

\begin{algorithm2e}[t]
  \SetAlgoNoLine
  \DontPrintSemicolon
  \For{\textrm{each partition} $q \in \mathcal{T}$}{ %
    $W(q) = \sum_{w_i \in w_q} \left( \sum_j w^{v_j}_i + \sum_k w^{e_k}_i \right)_{w_i}$\;
  }
  $W_{\textrm{max}} \longleftarrow ComputeGlobalMax(W)$\; 
  $W_{\textrm{avg}} \longleftarrow ComputeGlobalAverage(W)$\;
  \If{$W_{\textrm{max}} / W_{\textrm{avg}} > \kappa$}{
    $\mathcal{T'} \longleftarrow ComputeNewPartitions(\mathcal{T})$\; 
    $S \longleftarrow ConstructMatrix(\mathcal{T'})$\; 
    $\mathcal{G} \longleftarrow SolveMWBG(\mathcal{T'})$\; 
    $\mathcal{T} \longleftarrow RedistributeData(\mathcal{G})$\;

  }
  \caption{\label{alg:dyn} Dynamic load balancing framework.}
\end{algorithm2e}

\subsection{Performance Analysis}
 \label{sec:perf}

To evaluate the performance of the load balancer, we used \textsc{Cube} to solve two different incompressible flow problems on the K computer. And, the total execution time for performing a fixed number of time steps for both an unbalanced (no load balancing) and a balanced case (using load balancing) on various numbers of cores are compared. For both problems we used the QUICK scheme for the convective terms and a geometric multigrid solver for the pressure. Time integration was performed using a second order Crank--Nicholson method. The constraint based immersed boundary method, presented in this work, was used to represent the complex geometries (Fig. \ref{fig:cervo} $\&$ Fig. \ref{fig:pdcc}) used in the present analysis.

The geometries for the numerical experiments, namely the landing gear and the vehicle, were chosen to represent different types of immersed bodies over which  $\gamma$ can be studied. Although both the geometries are relatively complex, the overall particle density due of the two geometries are different. The particle density of the vehicle geometry is $2.68$ particles for every $100$ Eulerian cells, while that for the landing gear is $1.16$ particles. Here, particle density is defined as the ratio of the number of Lagrangian particles and the number of Eulerian cells.  Carrying out the numerical experiments on these two geometries will be helpful in understanding how the cost parameter varies across a range of immersed bodies.

% In the load balancer, the weights were assigned as described in Section \ref{sec:work}, with the additional immersed boundary cost added to $w_i^{v_{2}}$, modeled as a $\gamma \cdot n_\textrm{particles}$, with $\gamma$ a cost parameter usually in the range of $1-4$.

\subsubsection{Nose Landing Gear}
\begin{figure}[th]
  \begin{center}
    \subfloat[The landing gear geometry.]{\label{fig:pdcc}\includegraphics[width=0.49\columnwidth]{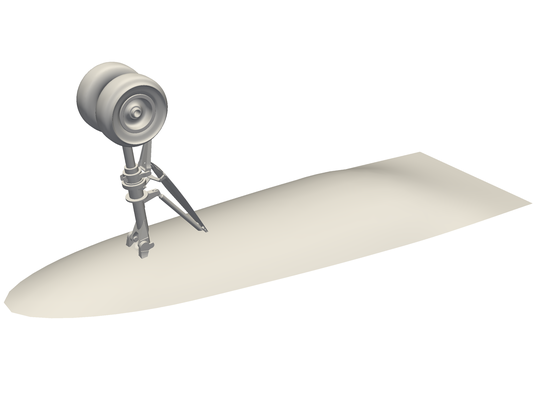}}
    \subfloat[The full vehicle model.]{\label{fig:cervo}\includegraphics[width=0.49\columnwidth]{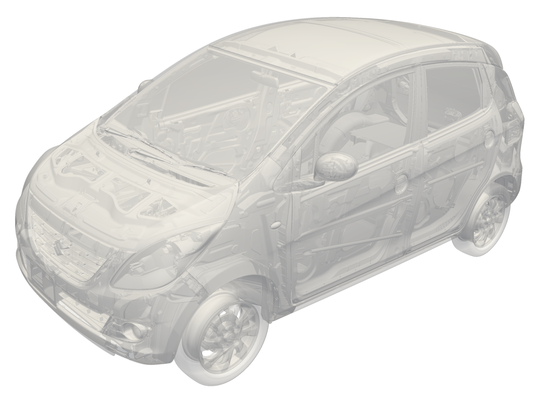}}
    \caption{Geometries used during performance analysis.}
  \end{center}
\end{figure}

The first problem is based on the nose landing gear (Fig. \ref{fig:pdcc}) case from AIAA's BANC series of benchmark problems. Our setup uses a mesh consisting of 48255 cubes, subdivided into $16$ cells in each axial direction, and the landing gear consists of $\approx0.5M$ surface triangles, resulting in $\approx 2.3M$ particles. %2279970 particles

% \begin{figure}[h]
%  \begin{center}
%    \subfloat[Unbalanced (Z-ordering).]{\label{fig:pdcc-unbalanced}\includegraphics[width=0.3\columnwidth]{pdcc_zorder_crop}}
%    \hspace*{1mm}
%    \subfloat[Balanced wrt. geometry.]{\label{fig:pdcc-balanced}\includegraphics[width=0.3\columnwidth]{pdcc_metis_crop}}
%  \end{center}
%   \caption{\label{fig:loadb} Load balancing wrt. immersed geometry and fluid cells, colored by MPI rank.}
% \end{figure}

\begin{figure}
  \begin{center}
    \includegraphics[width=0.49\columnwidth]{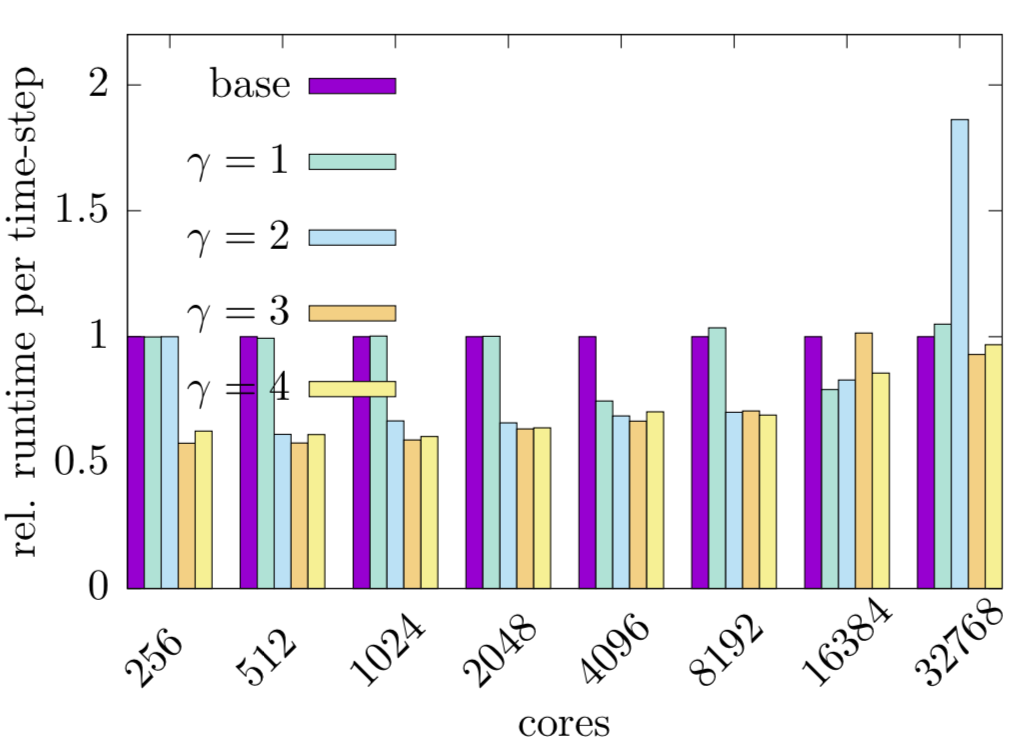}
%   \subfloat{\input{data/pdcc_k.tex}}
%    \subfloat{\input{data/pdcc_k_sp.tex}}
  \end{center}
%  \subfloat[Runtime per time-step.]{\label{fig:pdcctime}\input{data/pdcc_k.tex}}
%  \hspace*{0.1\textwidth}
%  \subfloat[Estimated workload imbalance.]{\label{fig:pdccloadb}\input{data/pdcc_k_balance.tex}}
%  \end{center}
  \caption{Realtive runtime per time-step for the load balanced and unbalanced cases for the nose landing gear.} % \rbcom{Remove runtime per time-step with estimated imbalance For cervo as well!The middle finger at 32768 on the right panel overly exagerates the $\gamma=2 $ case. We should perhaps replace the right panel with estimated imbalance?}}
  \label{fig:pdcctime}
\end{figure}

\begin{figure}
\centering
\includegraphics[width=0.49\columnwidth]{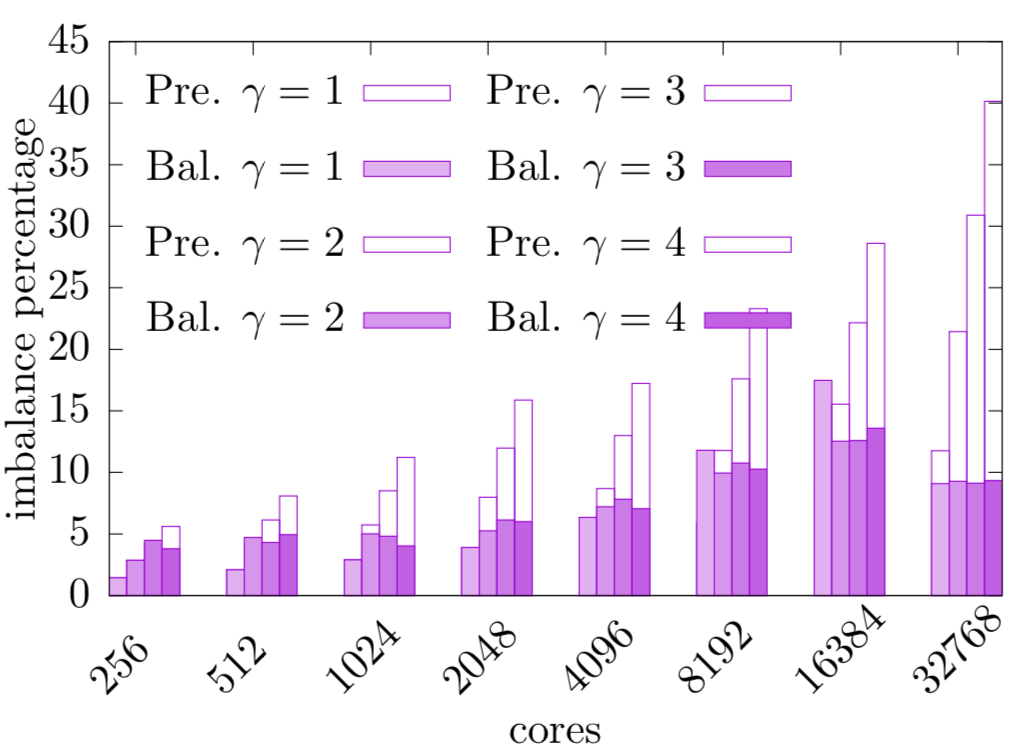}
\caption{Comparison of estimated workload imbalance before and after load balancing for the landing gear geometry. The imbalance data before the load balancing is shown in bars filled with white shade and labeled 'Pre.', and imbalance data after load balancing is shown in bars shaded in non white colors and labeled 'Bal.'  }
\label{fig:pdccimb}
\end{figure}

% \begin{table}[h]
% \begin{center}
%   \begin{tabular}{rrrrrr}
%     \hline
%     cores & Z-ordering & \multicolumn{4}{c}{Load Balanced} \\
%     & & $\gamma = 1$ & $\gamma = 2$ & $\gamma = 3$ & $\gamma = 4$ \\
%     \hline
% 256&     1107.408&        1106.028&        1106.841&        638.874&         692.552\\
% 512&     597.030&         593.504&         365.562&         344.813&         364.941\\
% 1024&    310.350&         311.224&         206.417&         183.070&         187.317\\
% 2048&   179.670&         179.930&         118.160&         113.850&         114.630\\
% 4096&    105.000&         78.136&          71.970&          69.794&          73.684\\
% 8192&    71.360&          73.873&          49.926&          50.300&          49.135\\
% 16384&   51.720&          40.864&          42.800&          52.470&          44.235\\
% 32768&   36.273&          38.064&          67.546&          33.710&          35.128\\
% \hline
%   \end{tabular}
% \caption{\label{tab:pdcc} Runtime per time-step for nose landing gear.}
% \end{center}
% \end{table}

% \begin{figure*}[t]
%  \begin{center}
%    \subfloat[Runtime per time-step.]{\label{fig:cervotime}\input{data/cervo_k.tex}}
%    \hspace*{0.1\textwidth}
%    \subfloat[Estimated workload imbalance.]{\label{fig:cervoloadb}\input{data/cervo_k_balance.tex}}
%  \end{center}
%  \caption{Runtime per time-step and estimated workload imbalance for the full car model.}
% \end{figure*}
In Fig. \ref{fig:pdcctime}  we present the relative time required to perform one time-step. We define relative runtime per time-step (RRPT) as the runtime per time-step (RPT) of a simulation normalized by runtime per time-step of unbalance base case, for each core count. The RRPT measure is helpful in gauging improvement (or the lack of it) of a load balanced simulation with respect to an unbalanced base case. From the results in Fig. \ref{fig:pdcctime} we can observe that by using the load balancer the runtime can be reduced to as much as $\approx 60\%$ of the unbalanced case. As the number of cores are increased, the gains of load balancing is less.

% This is most likely due to the fact that for a model with small particle density such as the landing gear, and few cubes in the mesh, the initial data decomposition will (for large core counts) already place several partitions close to the geometry, and indirectly balance the workload automatically.

%A similar observation can also be made in Fig. \ref{fig:pdccloadb}, where we present the estimated imbalance (Eq. \ref{eq:imb}) before and after load balancing. For a small amount of cores the estimated workload is greatly improved, while for the 8192 core case the effect of load balancing is less, due to more initial partitions.

The load balancing, or more specifically the estimated workload is also sensitive to the parameter $\gamma$ which controls how much more expensive it is to perform Lagrangian particle related operations compared to Eulerian operations. We can see that for $\gamma=1$, the lack of runtime improvement in the balanced case is independent of the core count. Except of the improvement at $4096$ and $16384$ cores, where the balanced case is slightly faster, the balanced and the unbalanced cases consume similar time per time-step. When $\gamma=2$, the balanced case outperforms the unbalanced case at all the core counts except at $256$ and $32768$ cores, while at $256$ the runtime for balanced and unbalanced is almost same, at $32768$ the balanced case is slower than the unbalanced case. Unlike $\gamma = 1\; \& \;\gamma =2$, the remaining two cost parameters, $\gamma = 3\; \& \;\gamma =4$, the balanced cases outperform the unbalanced case for all the core counts.

The estimated workload imbalance before and after the use of the load balancer is shown in Fig. \ref{fig:pdccimb}. The estimated imbalance before the load balancing is termed predicted imbalance, and the estimated imbalance after load balancing is called actual imbalance. The first observation that can be made is that, with the exception of $\gamma=1$, the predicted imbalance is typically greater than the actual imbalance when the load balancer is invoked. The load balancer is invoked when the predicted imbalance is greater than the threshold value of $4\%$, i.e. $\kappa=1.04$. When $\gamma=1$, for core count up to 2048, the predicted imbalance is less than $4\%$. As a result, the load balancer is not invoked, thus the actual imbalance and predicted imbalance are equal.

Next, we find that there are a few cases in which the predicted imbalance is less than the actual imbalance, yet the RRPT of these balanced cases is smaller than the base case. In other words, despite there being no reduction in the actual imbalance compared to the predicted imbalance, we see a speed-up in the runtime of the balanced case. Examples of these cases are ($\gamma=3$, core $=256$), ($\gamma=2$, $=512$), and ($\gamma=1$, core $=4096$) in Figs\ref{fig:pdcctime}~\&\ref{fig:pdccimb}. The likely reason for this unusual behavior is our choice of the definition of workload imbalance, $W_{\textrm{max}} / W_{\textrm{avg}} > \kappa$.  This definition of workload imbalance relies entirely on relative difference between maximum workload ($W_{\textrm{max}}$) and average workload ($W_{\textrm{avg}}$) and makes no consideration of the standard deviation of the workload of the system. For distributions with large standard deviation in workload distribution prior to load balancing, there is a large gap between maximum workload and minimum workload. And, there could be a large difference between maximum workload and median workload. For such a distribution there is scope for workload optimization through load balancing and overall speed-up in runtime. The data redistribution algorithm used in the load balancer always results in a `balanced-distribution' whose standard deviation in workload is small. But, it does not guarantee that maximum workload will be close to the average. As a result, there could be situations where the standard deviation in the workload has been reduced by the load balancer, but the maximum workload has not changed much (or may even have increased). Because the standard deviation has reduced, there could be speed-up in the runtime of the balanced case yet the actual imbalance based on our definition of imbalance may remain unchanged.

Lastly, another interesting observation we can make from the imbalance data in Fig. \ref{fig:pdccimb} is the trend of the predicted imbalance. The predicted imbalance progressively increases with core count. This trend implies that there is much scope for improvement the runtime of the base case through load balancing as the core count is increased. Yet, in Fig. \ref{fig:pdcctime} we see that the speed-up in runtime for the balanced case progressively decreases with increasing core count. This counter intuitive behavior is likely due to the lack of workload model for Lagrangian domain particle communication for imbalance estimation and for data redistribution by the load balancer. When the core count is low, overall number of partitions cutting the Lagrangian domain and the resulting communication cost would be small compared computational cost of Lagrangian operations. As the core count is increased, progressively more and more partitions cut the Lagrangian domain increasing the communication cost while the cost of Lagrangian computations on each partition decreases, which in turn increases relative cost of the Lagrangian communication. Modeling the workload due to communication of Lagrangian particles is important when core count is high. Thus, when the core count is high, where cost of  communication dominates (Lagrangian or otherwise), data redistribution by the current load balancer can result in a distribution whose communication pattern may not be very different from that of the unbalanced distribution. As a result, the runtime, which is now dominated by communication costs, of the balanced case may remain unchanged because the workload of Lagrangian communication was not modeled.

\subsubsection{Full Vehicle Model}
We consider the flow past a full vehicle model (see Fig. \ref{fig:cervo}) to evaluate the performance of the load balancer for geometries that result in higher particle density. The numerical methods used for this problem are identical to the ones used for the landing gear benchmark. We use a mesh consisting of 38306 cubes with $16^3$ cells per cube, and with a vehicle model consisting of $\approx12.5M$ surface triangles, resulting in $\approx 4.2M$ particles. % 4166058 \rbcom{particle number.}

% \begin{figure}[h]
%   \begin{center}
%     \includegraphics[width=0.5\columnwidth]{cervo}
%     \caption{\label{fig:cervo} The full car model.}
%   \end{center}
% \end{figure}

\begin{figure}
  \centering
%     \subfloat{\input{data/cervo_k.tex}}
  %   \subfloat{\input{data/cervo_k_sp.tex}}
  \includegraphics[width=0.49\columnwidth]{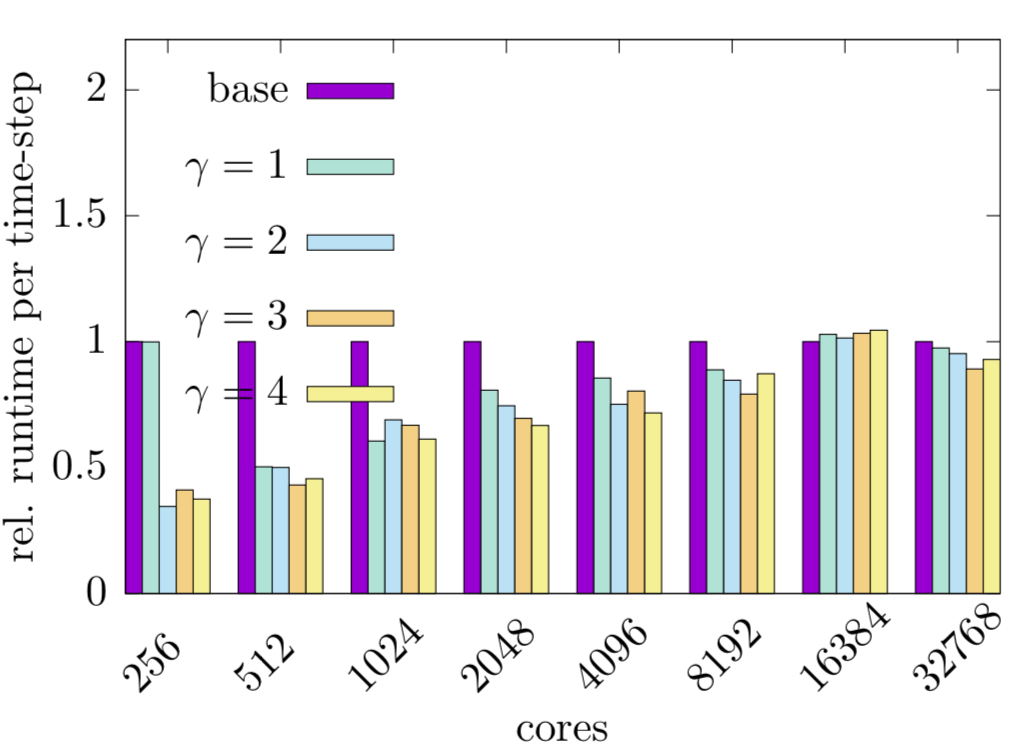}
  \caption{A comparison of  relative runtime per time-step of the load balanced and unbalanced cases of the full vehicle simulation at various core counts.}
  \label{fig:cervotime}
\end{figure}

Relative runtime per time-step data are presented in Fig. \ref{fig:cervotime}. It is seen that for all the chosen values of $\gamma$, $1-4$, the load balanced cases perform better than the unbalanced case.  For the cases with $256$ cores, the runtime reduced to $\approx 40\%$ or less of the unbalanced case when $\gamma$ was set to 2, 3 or 4. This improvement in the runtime gradually decreases with increasing core count. Unlike the landing gear case, in which a consistent improvement in runtime was seen only for $\gamma=3$ and $\gamma=4$, an improvement is seen for all values of $\gamma$ for the vehicle case. Although the trend may be a subtle one, it can seen that $\gamma=3$ and $\gamma=4$ more consistently result in a faster runtime compared to the other values of $\gamma$. Thus, with regards to the optimal choice of $\gamma$, based on the trend observed for the landing gear and the vehicle model cases, $\gamma=3$ or $\gamma=4$ are reasonable first approximations.

\begin{figure}
\centering
  \includegraphics[width=0.49\columnwidth]{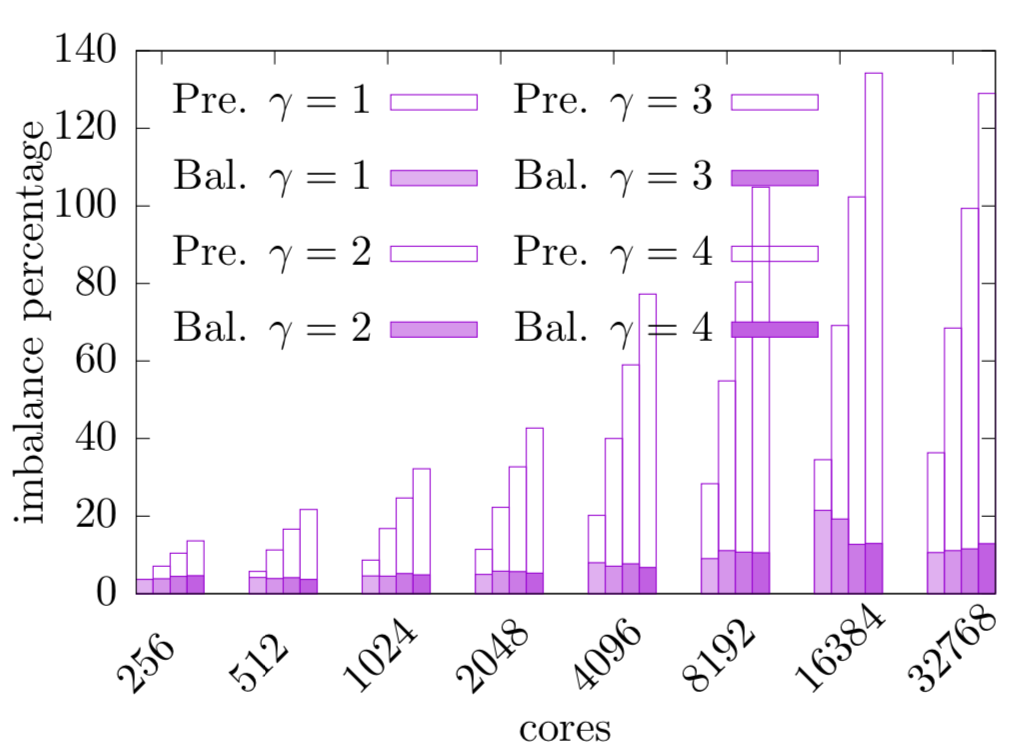}
\caption{Comparison of estimated workload imbalance before and after load balancing for the landing gear geometry. The imbalance data before the load balancing is shown in bars filled with white shade and labeled 'Pre.', and imbalance data after load balancing is shown in bars shaded in non white colors and labeled 'Bal.'.}
\label{fig:cervoimb}
\end{figure}

The estimated workload imbalance data for the full vehicle case is presented in Fig. \ref{fig:cervoimb}. It can be seen that the actual imbalance is consistently smaller than predicted imbalance for every all $\gamma$ and core count. This unlike the landing gear case in which we found that for a few case, the actual imbalance was greater than the predicted imbalance. This difference in trend, with respect to predicted and actual imbalance, between vehicle and landing gear could be due to particle density and volumetric distribution of particles. The particle density, of landing gear is 0.0116 and that of the full vehicle is 0.0268. The particle density of the vehicle is $\approx$~2.5 times that of the landing gear case. Owing to a greater volumetric space occupied by the vehicle model (as opposed to the smaller spatial volume occupied by the landing gear), there is greater volumetric spread of the particles relative to the mesh. A consequence of these two factors could be a workload distribution with a small standard deviation from the average workload which ensures that the actual imbalance is always smaller that the predicted imbalance. The trend of predicted imbalance, for all $\gamma$, with core count is identical to that of the landing gear case. The reason for this trend is already discussed in detail in the previous subsection.

% An improvement in runtime is observed up to a core count of $8192$, after which very little or no improvement is seen. For the nose landing gear on the other hand, the improvement in runtime persisted up to the highest core count case. This is most likely due to larger amount of particles   of the vehicle model relative to the mesh size in combination with the volumetric distribution of the particles in the mesh. Owing to a greater volumetric space occupied by the vehicle model (as opposed to the smaller spatial volume occupied by the landing gear),as the core count is increased, the original unbalanced partitioning will result in more MPI ranks around the vehicle, indirectly balancing the mesh partition. In contrast, because the landing gear occupies a smaller spatial volume, increasing core count does not guarantee more MPI ranks around the geometry. Consequently, there may still be scope for load balancing.

% \begin{table}[h]
% \begin{center}
%   \begin{tabular}{rrrrrr}
%     \hline
%     cores & Z-ordering & \multicolumn{4}{c}{Load Balanced} \\
%     & & $\gamma = 1$ & $\gamma = 2$ & $\gamma = 3$ & $\gamma = 4$ \\
%     \hline
% 256&     2053.329&        2052.014&        710.000&         843.484&         769.079\\
% 512&     953.600&         479.490&         477.416&       410.716&         434.653\\
% 1024&    406.230&        246.036&         280.220&       271.490&         249.00\\
% 2048&    206.130&        166.450&         153.685&       143.450&         137.505\\
% 4096&    122.330&        104.610&         91.889&        98.350&         87.664\\
% 8192&    81.750&         72.616&          69.249&        64.750&          71.360\\
% 16384&   57.413&          59.062&          58.229&        59.334&          60.010\\
% 32768&   54.870&         53.478&          52.276&        48.931&          51.000\\
% \hline
%   \end{tabular}
% \caption{\label{tab:cervo} Runtime per time-step for the full vehicle simulation.}
% \end{center}
% \end{table}

\section{Parallel Scalability}
\label{sec:scala}
The parallel scalability and computational efficiency of a software environment depends on the numerical algorithms involved as well as the design and implementation of each component of the software. In previous sections we have discussed the various components of the software, namely halo exchange, overlapped time-stepping, Lagrangian-Eulerian framework, etc., that we have attempted to optimize. While the gain through the optimization of individual components may not be necessarily high, the small gain in each of the components contributes to the overall computational efficiency and parallel scalability of the software. In this section we present results of strong scaling analysis of \textsc{Cube} with and without immersed geometries. In order to do this, we carried out simulations of flow around a Full-vehicle (Fig. \ref{fig:cervo}). The vehicle geometry was chosen because it is relatively large, dense and has a complex geometry, representative of geometries likely to be used in large scale simulations. Processor cores ranging from $1024$ to $65536$ were used to carry out the simulations to analyze the cost of the immersed body, and  the scalability of \textsc{Cube}. The mesh used for these simulations had $131,072$ cubes, with $16^3$ cells in each cube. QUICK scheme was used for the convection term and a geometric multigrid solver for the pressure. The maximum number iterations for the iterative conjugate gradient solver on the coarsest level was fixed at $10$ and the maximum number of v cycles were set to $50$. A motion was imposed on the vehicle by setting the axial velocity of the vehicle to $1$m/s. The magnitude of the vehicle's velocity is inconsequential because for any non-zero velocity of the vehicle, the Lagrangian particle MPI communicator is initiated. The particle communicator is not necessary when the vehicle is stationary. A non-stationary configuration of the vehicle is chosen for the strong scaling analysis because it is the computationally more expensive and severe IB simulation configuration.

Two sets of simulations were carried out, first without any immersed geometry and the second with the immersed geometry. These two sets of simulations will enable us to compare the strong scaling of the \textsc{Cube} with and without the immersed geometry.  The chosen geometry and mesh resulted in $5,097,567$ Lagrangian particles. In this subsection, we shall refer to simulations without immersed geometry as `channel' simulations (implying channel flow simulations), and the one with the immersed geometry as the `IB' simulation. In order to trigger the pressure solver in the channel flow simulation a no-slip boundary condition was imposed on the computational domain walls in the y direction. Slip boundary conditions were used along z-direction, and inflow-outflow boundary condition along x-direction. The same set of boundary conditions were used for the simulations with the immersed geometry.

\begin{figure*}[t]
 \begin{center}
%    \subfloat[]{\label{fig:SS}\includegraphics[width=0.45\columnwidth]{Speed_Up_CUBEvsIB.pdf}}
%    \hspace*{2mm}
   \centering
   \null\hfill
   \subfloat[Strong scalability]{\label{fig:SS}\includegraphics[width=0.49\columnwidth]{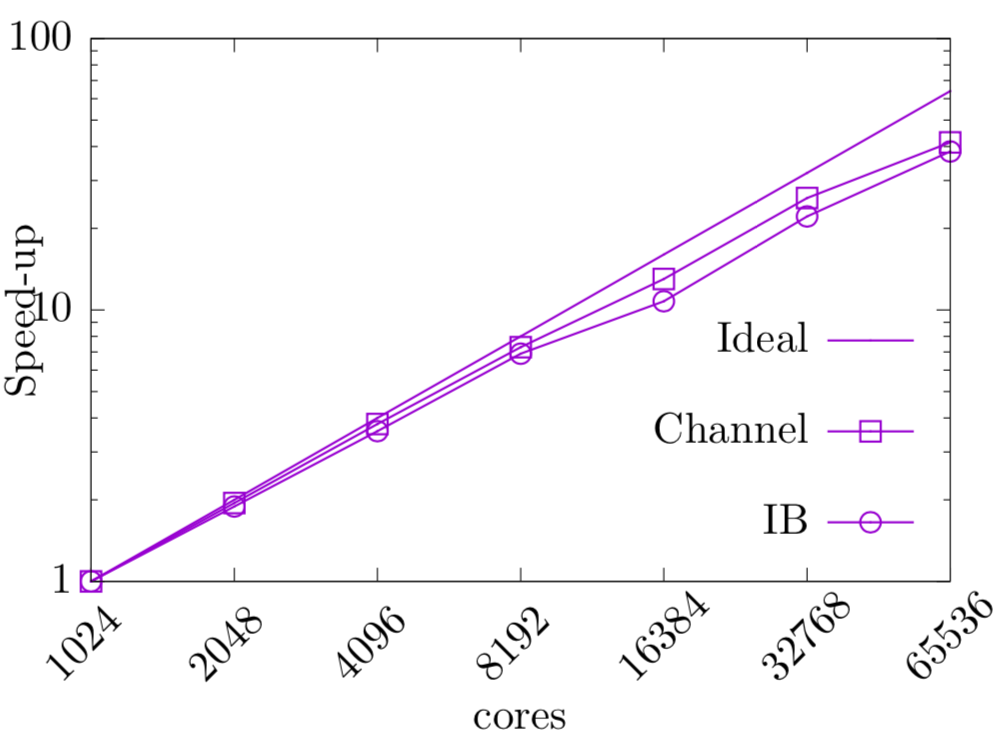}}
   \hfill
   \subfloat[Relative runtime]{\label{fig:SS-IB}\includegraphics[width=0.49\columnwidth]{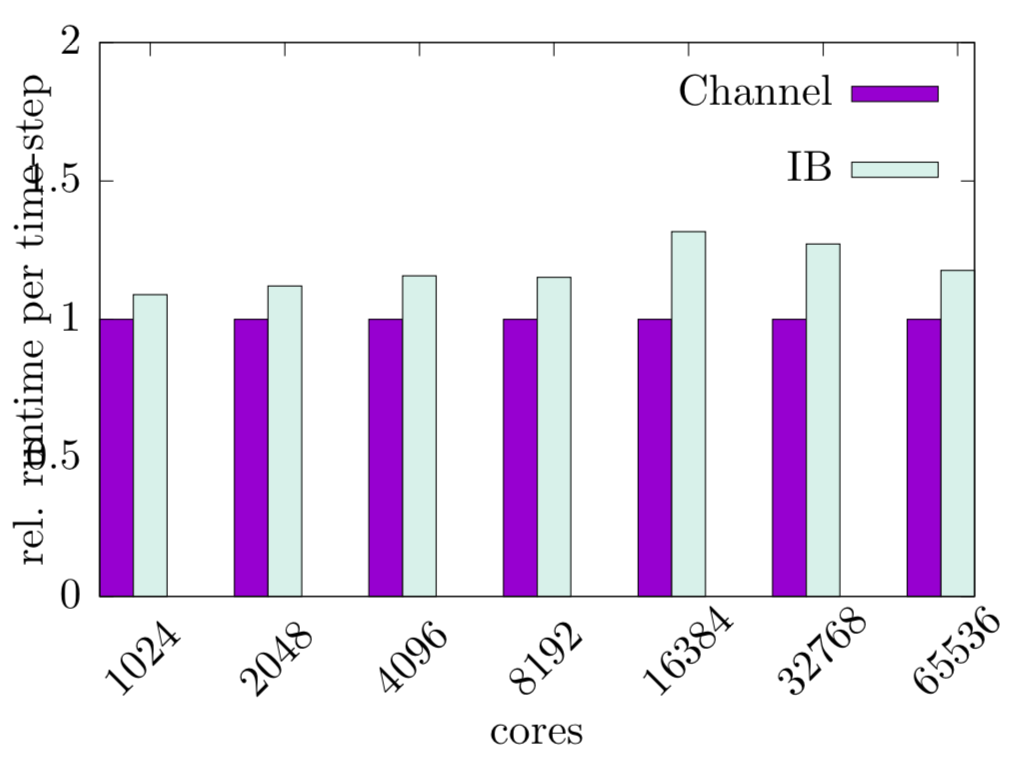}}
   \hfill\null
 \end{center}
  \caption{Strong scaling of \textsc{Cube} with and without an immersed geometry \protect\subref{fig:SS}. Comparison of the relative runtime per time-step of IB simulation with that of a corresponding Channel simulation \protect\subref{fig:SS-IB}. The runtime per time-step of each core cound case is normalized by that of the corresponding Channel case.}
  \label{fig:SS-CUBE}
\end{figure*}

The strong scaling results of the channel flow simulations and the simulations with moving immersed geometry are shown in Fig. \ref{fig:SS-CUBE}. Where, the speed up of cube is compared with the ideal speed up compared to the run-time of base case ($1024$ cores case). \textsc{Cube} shows good scalability and parallel efficiency all the way up to $65536$ cores for the both channel and IB cases. Furthermore, both channel and IB show very similar speed-up trend with the core count, which implies that the underlying scalability of the solver is not strongly influenced by the introduction of a geometry into the solver. This can also be seen in Fig. \ref{fig:SS-CUBE}b, which is described next.  The relative run-time per time-step is plotted in  Fig. \ref{fig:SS-CUBE}b. We define relative run-time per time-step as the run-time per time-step (RPT) of a simulation scaled by that of the corresponding channel simulation. For example, for any core count, say 4096, RPT of the IB simulation as well as that of channel simulation is normalized by the RPT of channel simulation of the 4096 cores case. This means that relative RPT will always be 1 for the channel case, as seen in Fig. \ref{fig:SS-CUBE}b. We can make two observations from this plot. First, scalability of \textsc{Cube} is not strongly influenced by the inclusion of an immersed geometry in the solver. Second, the plot reveals the relative cost of the inclusion of a complex geometry such as a full vehicle geometry. We find that the additional cost of inclusion of the IB varies between 10$\%$ and 25$\%$, which on average is around 15$\%$ which makes the simulation of large-scale industrial applications in \textsc{Cube} viable.
%
% For the channel flow case, \textsc{Cube} shows very good parallel efficiency upto to $32768$, for which the parallel efficiency is around $80\%$. The parallel efficiency drops from $80\%$ to $65\%$ when core count is increased to $65536$. This drop in parallel efficiency is likely due to saturation of the number of cubes per cores. For the largest case, there are only 2 cubes per core. In case of the simulations with immersed geometries, the trend and behavior of scalability and parallel efficiency is similar to that of channel flow case, although the parallel efficiency is slightly lower. A parallel efficiency of $85\%$ is observed upto to $8192$ core after which it drops to around $65\%$ and hovers around this value all the way up to largest core count case.

% \begin{figure}[tb]
%  \begin{center}
%   \includegraphics[width=0.5\columnwidth]{Parallel_eff_CUBEvsIB.pdf}
%  \end{center}
%   \caption{\label{fig:eta-CUBE} Parallel efficiency of \textsc{Cube} with and without immersed geometry.}
% \end{figure}

\section{Industrial applications}
\label{sec:IA}
\subsection{Flow around a vehicle in turning maneuver}
In this section we proceed to demonstrate the capability of \textsc{Cube} to simulate complex geometries undergoing complex motion. For this demonstration, we consider the flow around a vehicle in a turning maneuver. For this case a full-vehicle ``dirty'' CAD geometry is used. The CAD geometry is not subjected any special surface treatment to fix surface imperfections. All the intricate details of the vehicle geometry, except passenger seats and steering wheel, are retained. The passenger seats were removed to avoid generation of fine mesh inside the vehicle where the flow is not of interest.

The aerodynamic performance and stability a vehicle is strongly influenced by the crosswinds during cruise and while in turning maneuvers \cite{coop07,watk07}. It is difficult to simulate such real-world flow scenarios in wind tunnel experiments \cite{carl07,mack96}. Furthermore, it is also difficult to measure unsteady aerodynamic forces in wind tunnel experiments. Thus, it is desirable for numerical methods and frameworks to be able to efficiently and accurately simulate such flow conditions. To this end, here, we present simulation of a vehicle undergoing a turning motion, including wheel rotation and steer, chassis roll, turn and pitch, in a uniform flow.

The geometric extents of the vehicle are as follows: wheel base $\sim~2.4$ m, ground height $\sim~1.5$ m, width in span-wise direction $\sim1.7$~m. The computational domain extended from $-25$~m to $75$~m in the axial direction ($x$), $-25$~m to $25$~m in the span-wise direction ($y$) and $0$ to $25$~m in the vertical direction ($z$). The mesh size on the finest level near the vehicle surface was $\Delta\mathbf{x}\mid_{l_{7}}=6.1$~mm.  The mesh used for the simulation is shown in Fig. \ref{fig:bcm-lag-scaling}a. Slip boundary condition was imposed on the boundaries along vertical and span-wise directions, and inflow, outflow boundary conditions at $x^{-}$ and $x^{+}$, respectively. The viscosity and density of the fluid were $1.82\times 10^{-5}$~kg/(s$\cdot$m) and $1$~kg/m$^{3}$, respectively.  The vehicle is assumed to be moving at $13.89$~m/s (50~kmph) against an oncoming wind of $28.78$~m/s (100~kmph). Thus, a uniform flow of $41.67$~m/s is specified as the inflow condition at the $x^{+}$ boundary. An outflow boundary condition used at the $x^{+}$ boundary. A slip boundary condition is used on the boundaries along $y\;\&\; z$ directions. Reynolds number based on the height of the vehicle is $3.435 \times 10^{6}$.

%  \begin{figure}
%  \centering
%  \subfloat[]{\label{fig:turnxy}\includegraphics[width=0.4\textwidth]{Car-XY-alpha.pdf}}
%  \subfloat[]{\label{fig:turnyz} \includegraphics[width=0.4\textwidth]{Car-YZ-alpha.pdf}}
%  \caption{\label{fig:turn_schem} Schematic of a vehicle in a turning maneuver. The schematic shows the motion of the chassis  and the wheel (only the front left wheel is shown). The chassis is subjected to angular motion along $x$ and $z$ axes, while the wheel rotates about $y$ and $z$ axes. }
% \end{figure}

\begin{figure*}[t]
\centering
\null\hfill
\subfloat[$t = 0.3$ s.]{\label{fig:cervo-turn-xy-a}\includegraphics[width=0.24\textwidth]{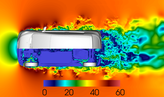}}
 \hfill
\subfloat[$t = 0.36$ s.]{\includegraphics[width=0.24\textwidth]{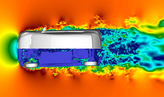}}
 \hfill
\subfloat[$t = 0.42$ s.]{\includegraphics[width=0.24\textwidth]{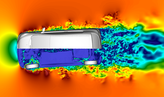}}
 \hfill
\subfloat[$t = 0.48$ s.]{\includegraphics[width=0.24\textwidth]{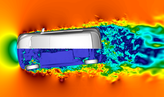}}
\hfill
\subfloat[$t = 0.54$ s.]{\includegraphics[width=0.24\textwidth]{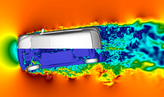}}
\hfill
\subfloat[$t = 0.60$ s.]{\includegraphics[width=0.24\textwidth]{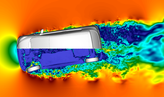}}
\hfill
\subfloat[$t = 0.66$ s.]{\includegraphics[width=0.24\textwidth]{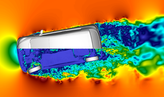}}
\hfill
\subfloat[$t = 0.72$ s.]{\label{fig:cervo-turn-xy-h}\includegraphics[width=0.24\textwidth]{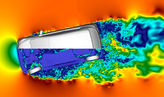}}
\hfill\null
\caption{Velocity magnitude visualized on an $xy$ plane at $z=0.25$~m from the simulation of flow around a vehicle in a turning maneuver.}
\label{fig:cervo-turn-xy}
\end{figure*}

\begin{figure*}
\centering
\null\hfill
\begin{minipage}{0.65\textwidth}
    \centering
    \subfloat[$t = 0.36$ s.]{\label{fig:cervo-turn-isosurf-a}\includegraphics[width=0.49\textwidth]{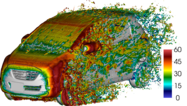}}
    \hfill
    \subfloat[$t = 0.48$ s.]{\includegraphics[width=0.49\textwidth]{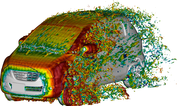}}
   \hfill
   \subfloat[$t = 0.60$ s.]{\includegraphics[width=0.49\textwidth]{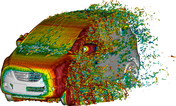}}   
   \hfill
   \subfloat[$t = 0.72$ s.]{\label{fig:cervo-turn-isosurf-d}\includegraphics[width=0.49\textwidth]{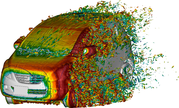}}
   \hfill
\end{minipage}
\begin{minipage}{0.34\textwidth}
  \begin{center}
    \subfloat[Aeorodynamic forces]{\label{fig:cervo-turn-forces}\includegraphics[width=\textwidth]{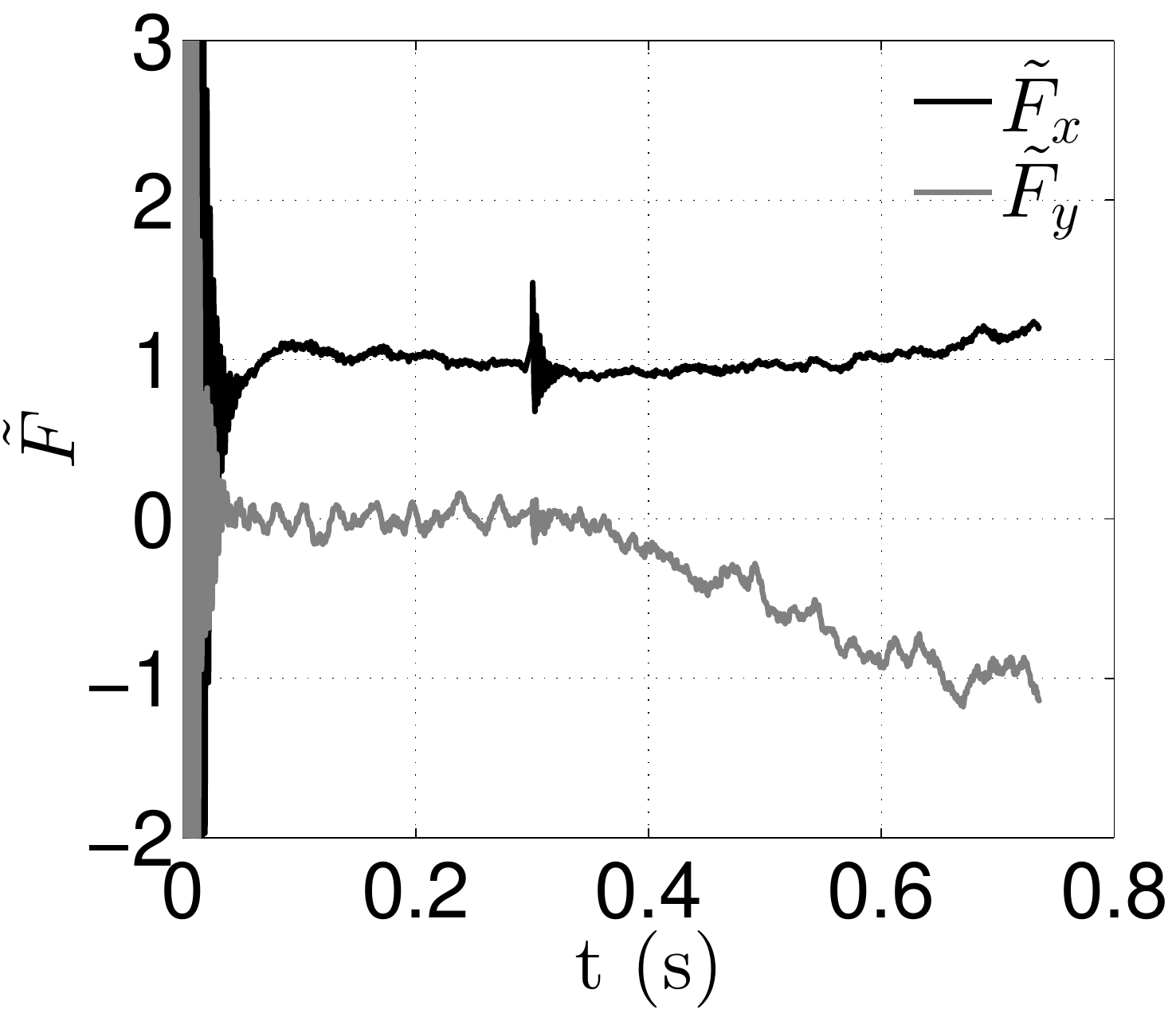}}
  \end{center}
\end{minipage}
% \hfill
\caption{Iso-surface of vorticity magnitude, coloured by velocity magnitude, of flow around a turning vehicle at different instants of the turn  \protect\subref{fig:cervo-turn-isosurf-a}--\protect\subref{fig:cervo-turn-isosurf-d}, and aerodynamic forces \protect\subref{fig:cervo-turn-forces}.}
\label{fig:cervo-turn-isosurf}
\end{figure*}

The vehicle motion is divided in time into two types of motion. First, from $t=0$~s to $t=0.3$~s the vehicle is assumed to be in a rectilinear motion, with a wheel rotation to match the vehicle speed of $13.89$~m/s. Next, for $t>0.3$~s the turning maneuver is imposed on the vehicle. A vehicle in a turning maneuver involves the following: Rotation of the chassis and wheels about the vertical axis ($z$-axis), roll of the chassis about the longitudinal axis ($x$-axis), additional rotation of the front wheels about the vertical axis which causes the vehicle turn. The wheels undergo rotation about the lateral axis ($y$-axis) when the vehicle is under motion relative to the fluid. The rotation of the vehicle and the wheels are imposed through angular velocities along respective directions. The angular velocity of the chassis is given by $\omega^{ch}=(\omega^{ch}_{x}, \omega^{ch}_{y} , \omega^{ch}_{z})\tanh\left(\alpha(t-0.3)\right)$,  the factor $\tanh\left(\alpha(t-0.3)\right)$ is used to gradually impose the rotation instead of an impulsively start and $\alpha$ governs the rate of change of the angular velocity. For our simulation we use $\omega^{ch}=(0.5, 0 , 0.025)\tanh\left(20(t-0.3)\right)$~rad/s. Similarly, $\omega^{wh}=(\omega^{wh}_{x}, \omega^{wh}_{y} , \omega^{wh}_{z})$ is the angular velocity of the wheels. The wheel rotation about the lateral axis is imposed impulsively at the start of the simulation ($t=0$~s) with $\omega^{wh}_{y} = 51.44$~rad/s. The angular velocity of the wheel turn of the front wheels is specified as $\omega^{wh}_{z} = 0.5\tanh\left(\alpha(t-0.3)\right)$. And, $\omega^{wh}_{x}=0$ for all the wheels and $\omega^{wh}_{z}=0$ for the rear wheels. It is to be noted that no rectilinear motion is directly imposed on the vehicle, linear motion of the vehicle is indirectly specified through the inflow condition which results in a relative velocity between the vehicle and the fluid that is equivalent to the relative velocity due to vehicle motion. Only rotational, angular velocities are directly imposed on the vehicle. This configuration is chosen to reduce size of the numerical mesh. By restricting the linear motion of the vehicle, the fine mesh is generated only in a small region around the vehicle. As the \textsc{Cube} does not have the capability to adaptively refine mesh as the immersed body moves, generating fine mesh around the vehicle path would result in an excessively large mesh.

% \begin{figure}[h]
%   \begin{center}
%     \includegraphics[width=0.4\columnwidth]{figs/cervo-turn/drag_data/Cervo-turn-forces.pdf}
%     \caption{\label{fig:cervo-turn-forces} The landing gear geometry.}
%   \end{center}
% \end{figure}

Visualization of the results from the simulation are presented in Figs. \ref{fig:cervo-turn-xy}~$\&$\ref{fig:cervo-turn-isosurf}.  The velocity magnitude on a horizontal plane ($xy$ plane) at $z=0.25$~m is plotted in Fig. \ref{fig:cervo-turn-xy}. During visualization, the vehicle chassis is box-sliced at $y=0$ such that chassis is visible only for $y>0$. This enables the visualization of the front left wheel's turning motion, as shown in  Fig. \ref{fig:cervo-turn-xy}. It can be seen in the figure that the vehicle and the wheel's turning begins at $t=0.3$~s and proceeds until the end of the simulation. As the wheels turn and lose alignment with the chassis (around $t=0.36$~s in Fig. \ref{fig:cervo-turn-xy}), a more pronounced separation flow  is caused around the wheels compared to relatively less separated flow around the chassis. This is more evident in Fig. \ref{fig:cervo-turn-isosurf-a} -- \ref{fig:cervo-turn-isosurf-d}, in which iso-surfaces of the vorticity magnitude is shown at four successive instants of the vehicle's turn. The reason for this more pronounced separation near the front left wheel is due to two likely causes. First, as the wheel turns, it looses alignment with the chassis and projects out. Second, interaction between a rotating wheel and the oncoming fluid could interact turbulently leading to the separation. In the sequence of images in Fig. \ref{fig:cervo-turn-xy} it can be seen that at all instances of the turn, the separation is actually induced by the wheel, not by front bumper.

The forces on the vehicle during the turning maneuver is plotted in Fig. \ref{fig:cervo-turn-forces}. Forces normalized by the mean axial force, $\tilde{F}=F/F^{avg}_{x}$, are plotted in the figure. The lateral force $F_{y}$ oscillates about a zero mean until the beginning of the turning motion after which an approximate linear increase with time is seen. The rate of change of the lateral force is as expected. but the lateral force approximately equal in magnitude to the axial force despite the small angle of turn. Furthermore, it is interesting to note that the rate of change of axial and lateral forces during the turn are very different. The rate of change of the axial force is very small and consequently increase in the force due to the turn is very small compared to the increase in the lateral force. Also, the axial force sees a more rapid increase towards the end of the turn (between $t=0.65$~s to $t=0.72$~s), which implies that the axial force is severely affected when the turn angle is greater than a certain critical value.  If the vehicle continues to turn in an axially dominant wind environment, the axial force could increase more rapidly affecting vehicle's stability.

% The relatively large lateral force due to cross wind type flow during turning could adversely affect the stability of the vehicle.
% The lateral force and the rise of the axial force towards the end of the turn in the simulation are indicative of forces the vehicle is likely to experience in real cross wind flows (where lateral flow velocity dominated axial flow velocity).

\section{Conclusions}
\label{sec:sum}
In this work we have presented \textsc{Cube}, a highly scalable framework for large-scale industrial simulations. A Lagrangian-Eulerian based constraint immersed boundary method adapted to the BCM framework was presented. The immersed boundary method is implemented such that it almost entirely eliminates pre-processing of input CAD geometries and allows that usage of `dirty' CAD data, which was demonstrated in Section\ref{sec:IA}. An Lagrangian data structure, the building cube Lagrangian, that enables efficient Lagrangian-Eulerian interpolation was presented. A spatial decomposition strategy of parallelization for the Lagrangian-Eulerian system in \textsc{Cube} was used in this work.

Various optimization strategies, overlapped halo exchange, overlapped time stepping, I/O strategies, incorporated in \textsc{Cube} were presented. A multi-threaded halo exchange method was presented and a closely related overlapped time-stepping scheme was also presented. Initial test of overlapped time stepping scheme showed a gain by as much as factor 1.6 in runtime. Through MPI I/O we were able to achieve a throughput of 10~GB/s for read performance on the K computer for file sizes of 251 and 618 GB. Through a wavelet based data compression algorithm, a compression ratio of 15 was achieved on \textsc{Cube}. A multi-constraint based load balancing framework for the Lagrangian-Eulerian system was presented. Upto 50$\%$ gain in runtime was achieved through the load balancer.  Despite the large gains in runtime for most cases, for some cases when core count was large, lack of workload modeling for the Lagrangian communication lead to a significant drop in the gain in speedup. Lastly, a strong scaling analysis of \textsc{Cube} was presented where we show good scalability upto  65536 cores.

Future work includes further optimization of communication patterns, especially on emerging hardware (e.g. massively parallel multicore processors), tuning and improved workload modeling in the load balancing framework, and further development of the numerical methods to extend the framework's multiphysics capabilities.

\section*{Acknowledgments}
% \begin{funding}
This research was supported by MEXT as “Priority Issue on Post-K computer” (Development of innovative design and production processes) and used computational resources of the K computer provided by the RIKEN Advanced Institute for Computational Science through the HPCI System Research project (Project ID: hp150284 and hp160232).

%This work was supported through the computing resources provided on the K computer by RIKEN Advanced Institute for Computational Science. %(proposal no. ra000012).
% \end{funding}

% \bibliographystyle{elsarticle-num}
% \bibliography{sisc17.bib} 
\end{document}